\documentclass[a4paper,12pt]{article} 
\usepackage[english]{babel}
\usepackage{pgfplots}
\usepackage[T1]{fontenc} 
\usepackage[utf8]{inputenc}
\usepackage{graphicx}
\usepackage{pgf,xcolor}
\usepackage{algpseudocode}
\usepackage{hyperref}

\usepackage{amssymb,amsmath}
\usepackage{algorithm}
\usepackage{caption}
\usepackage{graphics,psfrag}
\usepackage{bbm}
\usepackage{amsmath}

\makeatletter
\newcommand{\distas}[1]{\mathbin{\overset{#1}{\kern\z@\sim}}}%
\newsavebox{\mybox}\newsavebox{\mysim}
\newcommand{\distras}[1]{%
  \savebox{\mybox}{\hbox{\kern3pt$\scriptstyle#1$\kern3pt}}%
  \savebox{\mysim}{\hbox{$\sim$}}%
  \mathbin{\overset{#1}{\kern\z@\resizebox{\wd\mybox}{\ht\mysim}{$\sim$}}}%
}
\makeatother

\usepackage{natbib}
\usepackage{makeidx}
\usepackage{tikz}
\usetikzlibrary{matrix,arrows,shapes,positioning,mindmap,trees,automata}

\makeindex

\begin{document}

\title{\textbf{Bayesian Nonparametric Modelling of Conditional Multidimensional Dependence Structures}}
\author{Rosario Barone \textsuperscript{1}, Luciana Dalla Valle \textsuperscript{2}}
\date{ }


\maketitle
        \vspace{-40pt}
 \begin{center}
 {\small \textsuperscript{1}
        Sapienza University of Rome, Italy  \hspace{18pt}
        \textsuperscript{2}
             University of Plymouth, UK}
\end{center}   


\date{}

\def\spacingset#1{\renewcommand{\baselinestretch}%
{#1}\small\normalsize} \spacingset{1}



\bigskip

\begin{abstract}
In recent years, conditional copulas, that allow dependence between variables to vary according to the values of one or more covariates, have attracted increasing attention. In high dimension, vine copulas offer greater flexibility compared to multivariate copulas, since they are constructed using bivariate copulas as building blocks. In this paper we present a novel inferential approach for multivariate distributions, which combines the flexibility of vine constructions with the advantages of Bayesian nonparametrics, not requiring the specification of parametric families for each pair copula. Expressing multivariate copulas using vines allows us to easily account for covariate specifications driving the dependence between response variables. More precisely, we specify the vine copula density as an infinite mixture of Gaussian copulas, defining a Dirichlet process (DP) prior on the mixing measure, and we perform posterior inference via Markov chain Monte Carlo (MCMC) sampling. Our approach is successful as for clustering as well as for density estimation. We carry out intensive simulation studies and apply the proposed approach to investigate the impact of natural disasters on financial development. Our results show that the methodology is able to capture the heterogeneity in the dataset and to reveal different behaviours of different country clusters in relation to natural disasters. 

\end{abstract}

\noindent%
\vfill

\newpage
\spacingset{1.75} 
\section{Introduction}

In many real data applications we are often required to model jointly $d \geq 3$ continuous random variables, denoted as $Y_1,\dots,Y_d$. The multivariate distribution, which allows us to describe the joint behaviour of those variables, can be denoted as $F(Y_1,\dots,Y_d)=P(Y_1\le y_1,\dots,Y_d,\le y_d)$.
However, complex relations between data, particularly asymmetric and tail dependent associations, are often difficult to be modelled.
The copula approach allows us to express the multivariate distribution of a set of variables by separating the marginals from the dependence structure.
Copulas were introduced by \cite{sklar1959fonctions} and, since then, they have been applied in a wide variety of fields (see, for example, \cite{kolev2006copulas} for a review, and \cite{genest2009advent} and \cite{fan2014copulas} for applications in finance and economics, respectively).

In recent years, the idea of modelling the effect of covariates on the dependence structure described by copulas has attracted increasing attention.
\cite{patton2006modelling}, \cite{jondeau2006copula} and \cite{bartram2007euro}  considered time-varying dependence copula parameters in time series analysis.
\cite{acar2011dependence} estimated the functional relationship between copula parameters and covariates adopting a non-parametric approach.
\cite{craiu2012mixed} introduced a bivariate conditional copula model for continuous or mixed outcomes. 
\cite{abegaz2012semiparametric} and \cite{gijbels2011conditional}, respectively, suggested semiparametric and non-parametric approaches for the estimation of conditional copulas, proving the consistency and the asymptotic normality of the estimators.

For inference, several contributions in the literature follow the Bayesian nonparametric approach.
\cite{wu2014bayesian} presented a Bayesian nonparametric method for estimating multivariate copulas using a DP mixture of multivariate skew-Normal copulas, and \cite{wu2015bayesian} proposed a DP mixture of bivariate Gaussian copulas. In both cases the authors performed posterior inference via slice sampling \citep{walker2007sampling, kalli2011slice}.
\cite{valle2018bayesian} extended the approach of  \cite{wu2015bayesian} to bivariate conditional copulas, introducing dependence from covariates and implementing Bayesian nonparametric inference via an infinite mixture model. 

A different approach is followed by \cite{grazian2017approximate}, who described an approximate Bayesian inference method for semiparametric bivariate copulas, based on the empirical likelihood. 
This approach is extended by \cite{grazian2021approximate}, who compared
several Bayesian methods to approximate the posterior distribution of functionals of the dependence including covariates, using
nonparametric models which avoid the selection of the copula function.

Vines are multivariate
copulas constructed by using only bivariate building blocks which can be selected independently \citep{czado2019analyzing}.
This class of flexible
copula models has become very popular in the last years for many applications in diverse fields such as finance, insurance, hydrology, marketing, engineering, chemistry, aviation, climatology and health.
The popularity of vine copulas is due to the fact that they allow, in addition to the separation of margins and dependence by the copula approach, tail asymmetries and separate multivariate component modeling \citep{aas2009pair}.

In this paper we propose a Bayesian nonparametric inferential approach for multivariate copulas, which we express using vines.
We consider a DP mixture of vine copulas, assuming a DP prior distribution on the mixing measure of an infinite mixture of vine copulas. 
Bayesian nonparameteric inference for mixture copula models was adopted by \cite{zhuang2021bayesian} to group similar dependence structures. However, the authors restricted their attention to the single-parameter unconditional copula functions and the bivariate scenario.
Our approach has proven to be particularly successful in the conditional case, where the effects of covariates on the dependence structures  are considered. 
Here, the constraints on the correlation matrix to which traditional multivariate copulas are subjected to would make the specifications of covariates arduous.
Our approach overcomes the limitations of multivariate copula modelling, combining the flexibility of the vine construction with the advantages of the Bayesian nonparametric approach and allowing for both clustering and density estimation. 

Our method allows us to model the unobserved heterogeneity, which is often present in real datasets, in a natural and flexible way. 
Bayesian mixture models that account for heterogeneity were implemented for example by \cite{buddhavarapu2016modeling}, who incorporated heterogeneity in the model using a finite multivariate normal mixture prior on the random parameters.
\cite{zhang2018modeling} adopted a similar approach to capture heterogeneity in learning styles in a dataset collected from a computer-based learning system.
Following the Bayesian nonparameteric strand in the literature, \cite{green2001modelling} modelled heterogeneity with DP based models illustrated in the context of univariate mixtures. 
DP mixtures were also employed by \cite{turek2021bayesian} to detect heterogeneity in ecological datasets.
However, to the best our knowledge, this paper is the first to apply the Bayesian nonparametric approach to conditional vine copulas mixtures for analysing heterogeneous data.
We demonstrate that our method allows us to capture the unobserved heterogeneity in a real dataset that relates financial development in different countries to the occurrence of natural disasters.
Our approach identifies two distinct country clusters. In the first one, the financial development temporal dependence is negatively affected by natural calamities, while in the second one we observe a positive effect.
These results reflect government preparedness to face natural hazards.

The remainder of the paper is organised as follows. Section 2 introduces unconditional and conditional vine copulas; Section 3 illustrates the DP mixture approach for conditional vines; Section 4 focuses on the implementation of the model using the Gibbs sampling algorithm; Section 5 applied the proposed approach to simulated and real datasets; concluding remarks are given in Section 6. 



\section{Background and Preliminaries}\label{sec:background} 
 
\subsection{Vine Copulas}\label{sec:vines}

Let us consider the random variables $Y_1,\dots,Y_d$, whose joint distribution function is denoted by $F$, with margins $F_{1},\dots,F_{d}$. Then, there exists a copula distribution function $C$, with $C:[0,1]^{d} \to [0,1]$  such that 
\begin{equation} \label{eq:sklar}
F(y_1,\dots,y_d)=C(F_{1}(y_1),\dots,F_{d}(y_d)).
\end{equation}
If $F_{1},\dots,F_{d}$ are continuous, then $C$ is unique. 
Equation \eqref{eq:sklar} is known as Sklar's theorem \citep{sklar1959fonctions}, which is the most important result in copula theory.
The theorem states that multivariate distributions can be represented by the copula of its marginals, allowing the separation between the margins from the copula. 

Several applications of copula modelling to real data problems are in dimension $d=2$, where a wide range of parametric copula families exists \citep{nelsen2007introduction, joe2014dependence}.
On the other hand, a number of interesting data applications involve dimensions $d\ge3$.
In this case, multivariate copulas are the most common option.
However, two main drawbacks make multivariate copula modelling unattractive. First of all, there are fewer parametric families available in $d\ge3$ compared to the bivariate case; furthermore, the dependencies between each pair of variables are assumed to belong to the same parametric family. These reasons often make multivariate copula models an undesirable option.

The drawbacks of multivariate copulas were noticed by \cite{aas2009pair}, who proposed a wider class of multivariate copulas, based on the pair-copula construction method introduced by \cite{joe1996families} and later discussed by \cite{bedford2001probability} and \cite{kurowicka2006uncertainty}. The method consists in rewriting a $d$-variate copula as product of $d(d-1)/2$ bivariate copulas, that can be represented as a graphical model, called vine. This approach is more flexible than traditional multivariate copulas, as the bivariate pair-copulas constituting the vine can be selected from a wide range of parametric families \citep{czado2019analyzing}. 

Let us focus, as an example, on dimension $d=3$. In this case, according to the \eqref{eq:sklar}, the joint distribution of the random variables $Y_1$, $Y_2$ and $Y_3$ can be written as 
\begin{equation}\label{eq:sklar_3dim}
F(y_1,y_2,y_3)=C(F_{1}(y_1),F_{2}(y_2),F_{3}(y_3)).
\end{equation}
The \eqref{eq:sklar_3dim} can be expressed in terms of densities
\begin{equation}\label{eq:sklar_3dim_dens}
f(y_1,y_2,y_3)=c(F_{1}(y_1),F_{2}(y_2),F_{3}(y_3)) \prod_{j=1}^{3} f_j(y_j)
\end{equation}
where $f(y_1,y_2,y_3)$ is the joint density of $Y_1$, $Y_2$ and $Y_3$, $f_{1}(y_1)$, $f_{2}(y_2)$ and $f_{3}(y_3)$ are the marginal densities and $c(\cdot,\cdot,\cdot)$ is the corresponding trivariate copula density.
Now, it can be shown that the joint three dimensional density in the \eqref{eq:sklar_3dim_dens} can be expressed
in terms of bivariate copulas and conditional distribution functions
according to the following pair copula decomposition 
\begin{equation}\label{eq:ppc_3d}
\begin{split}
f(y_1,y_2,y_3) & =  c_{1,2}(F_{1}(y_1),F_{2}(y_2))  c_{2,3}(F_{2}(y_2),F_{3}(y_3))   \times \\
& c_{1,3;2}(F_{1|2}(y_1|y_2),F_{3|2}(y_3|y_2)) 
 f_{1}(y_1) f_{2}(y_2) f_{3}(y_3) 
\end{split}
\end{equation}
where $c_{1,3;2}(\cdot,\cdot; y_2)$ denotes the copula density associated with the conditional distribution of $(Y_1, Y_3)$ given $Y_2=y_2$, $F_{1|2}(y_1|y_2)$ is the conditional distribution of $Y_1$ given $Y_2$, $F_{3|2}(y_3|y_2)$ is the conditional distribution of $Y_3$ given $Y_2$, $c_{1,2}(\cdot,\cdot)$ is the copula density of $(Y_1, Y_2)$ and $c_{2,3}(\cdot,\cdot)$ is the copula density of $(Y_2, Y_3)$. 
According to \cite{joe1996families}, the conditional distribution functions in the \eqref{eq:ppc_3d} can be derived in terms of (derivatives of)
pair-copula components, such that, for example
\begin{equation*}
F_{1|2}(y_1|y_2)=\frac{\partial C_{1,2}(F_{1}(y_1),F_{2}(y_2))}{\partial F_{2}(y_2)}.
\end{equation*}

Pair copula constructions are generally not unique, since a different ordering of the variables typically generates a different pair copula decomposition.
In the \eqref{eq:ppc_3d} we considered $Y_2$ as the central variable, driving the dependence with $Y_1$ and $Y_3$.
In the practice, the variable ordering is typically established either from context or by preliminary estimation of the strength of the pairwise associations \citep{czado2019analyzing}.

\tikzstyle{observed} = [circle, draw, thin, minimum height=1.5em]
\begin{figure}[h]
\centering
\resizebox{0.5\textwidth}{!}{%
\begin{tikzpicture}[node distance=4cm, auto,>=latex', thick]
    \path[->] node[observed] (1) {$1$};
    \path[-] node[observed, below right of=1] (12) {$1,2$};
    \path[-] node[observed, above right of=12] (2) {$2$}
    				(1) edge node {$1,2$} (2);  
   	\path[-] node[observed, below right of=2] (23) {$2,3$}
                  (12) edge node {$1,3;2$} (23);
	\path[-] node[observed, above right of=23] (3) {$3$}
                  (2) edge node {$2,3$} (3);      
\end{tikzpicture}
}%
\caption{Trivariate vine representation.}\label{fig:simp_vine}
\end{figure}
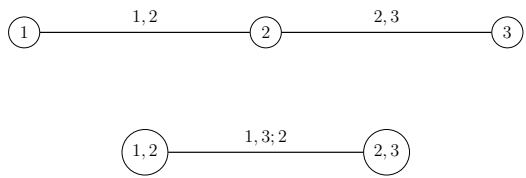

Pair copula constructions can be represented in graphical form via a vine representation. 
For example, Figure \ref{fig:simp_vine} depicts the trivariate vine which corresponds to the pair copula construction in \eqref{eq:ppc_3d}.
The different levels are called \textit{trees}, hence the trivariate vine in Figure \ref{fig:simp_vine} has two trees.
In higher dimension, the vine representation can be generalized to special vine distribution classes known as D- and C-vines (namely, drawable and canonical vines, two examples of which are shown in Figures \ref{fig:d_vine} and \ref{fig:c_vine} respectively) or, more generally, R-vines (regular vines) (see \cite{bedford2001probability},
\cite{aas2009pair} and \cite{czado2019analyzing}).

\tikzstyle{observed} = [circle, draw, thin, minimum height=1.5em]
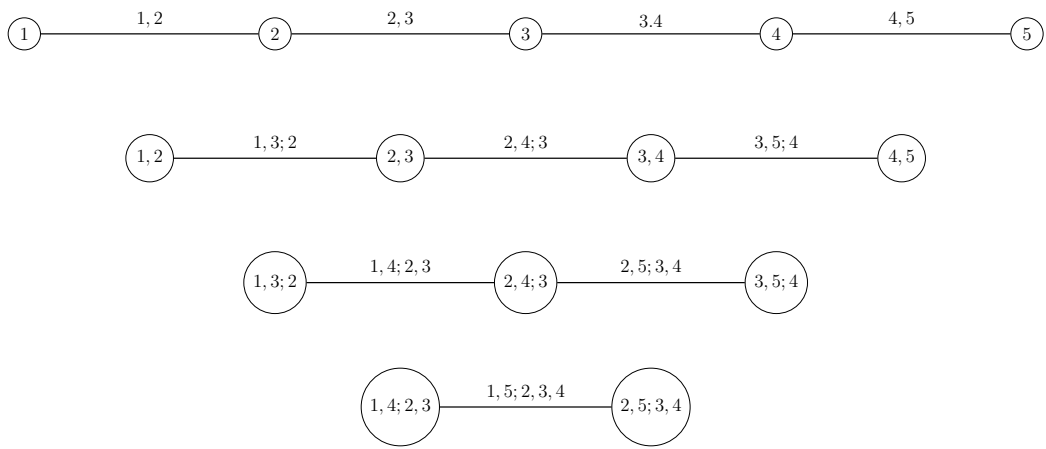
\begin{figure}[htbp]
\centering
\resizebox{1\textwidth}{!}{%
\begin{tikzpicture}[node distance=4cm, auto,>=latex', thick]
    \path[-] node[observed] (1) {$1$};
    \path[-] node[observed, below right of=1] (12) {$1,2$};
    \path[-] node[observed, above right of=12] (2) {$2$}
    				(1) edge node {$1,2$} (2);  
   	\path[-] node[observed, below right of=2] (23) {$2,3$}
                  (12) edge node {$1,3;2$} (23);
	\path[-] node[observed, above right of=23] (3) {$3$}
                  (2) edge node {$2,3$} (3);  
    \path[-] node[observed, below right of=3] (34) {$3,4$}
                  (23) edge node {$2,4;3$} (34);
	\path[-] node[observed, above right of=34] (4) {$4$}
                  (3) edge node {$3.4$} (4);   
    \path[-] node[observed, below right of=4] (45) {$4,5$}
                  (34) edge node {$3,5;4$} (45);
	\path[-] node[observed, above right of=45] (5) {$5$}
                  (4) edge node {$4,5$} (5);   
    \path[-] node[observed, below right of=12] (132) {$1,3;2$};  
    \path[-] node[observed, below right of=132] (1423) {$1,4;2,3$};
    \path[-] node[observed, above right of=1423] (243) {$2,4;3$} (132) edge node {$1,4;2,3$} (243);  
    \path[-] node[observed, below right of=243] (2534) {$2,5;3,4$} (1423) edge node {$1,5;2,3,4$} (2534);
     \path[-] node[observed, above right of=2534] (354) {$3,5;4$} (243) edge node {$2,5;3,4$} (354);  
\end{tikzpicture}
}%
\caption{D-vine tree sequence in $5$-dimensions and four trees.}\label{fig:d_vine}
\end{figure} 

\tikzstyle{observed} = [circle, draw, thin, minimum height=1.5em]
\begin{figure}[htbp]
\centering
\resizebox{0.4\textwidth}{!}{%
\begin{tikzpicture}[node distance=4cm, auto,>=latex', thick]
    \path[-] node[observed] (1) {$1$};
    \path[-] node[observed, above left of=1] (2) {$2$} 
    (1) edge node {$1,2$} (2);  
    \path[-] node[observed, above of=1] (3) {$3$} 
    (1) edge node {$1,3$} (3); 
     \path[-] node[observed, above right of=1] (4) {$4$} 
    (1) edge node {$1,4$} (4); 
   	  \path[-] node[observed, right of=1] (5) {$5$} 
    (1) edge node {$1,5$} (5); 
     \path[-] node[observed, below right of=1] (14) {$1,4$};
     \path[-] node[observed, below left of=14] (12) {$1,2$} (12) edge node {$2,4;1$} (14); 
     \path[-] node[observed, above of=12] (13) {$1,3$} (12) edge node {$2,3;1$} (13);
     \path[-] node[observed, right of=12] (15) {$1,5$} (12) edge node {$2,5;1$} (15);
     \path[-] node[observed, below right of=12] (241) {$2,4;1$};
     \path[-] node[observed, below left of=241] (231) {$2,3;1$} (231) edge node {$3,4;1,2$} (241);
     \path[-] node[observed, right of=231] (251) {$2,5;1$} (231) edge node {$3,5;1,2$} (251);
     \path[-] node[observed, below right of=231] (2512) {$3,5;1,2$};
     \path[-] node[observed, left of=2512] (3412) {$3,4;1,2$} (3412) edge node {$4,5;1,2,3$} (2512);
\end{tikzpicture}
}%
\caption{C-vine tree sequence in $5$-dimensions and four trees.}\label{fig:c_vine}
\end{figure}
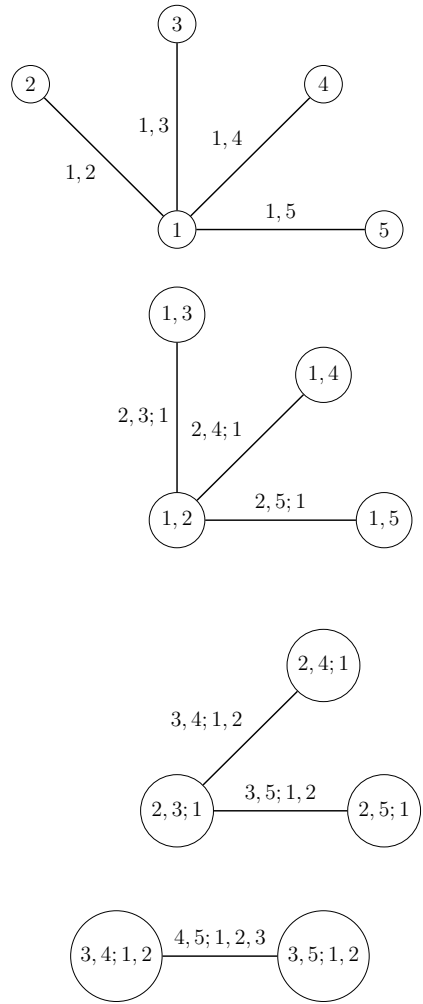 

\newpage
Given the random variables $Y_1,\dots,Y_d$, their joint density can be decomposed as a D-vine as follows
\begin{multline*}
f(y_1, \ldots, y_d) = \left[  
\prod_{\ell=1}^{d-1}\prod_{k=1}^{d-\ell}c_{k,\ell+k; k+1, \ldots, k+\ell-1}
\left\{
F_{k | k+1, \ldots, k+\ell-1} (y_k | y_{k+1, \ldots, k+\ell-1}),  \right. \right. \\
\left. \left. F_{\ell+k | k+1, \ldots, k+\ell-1} (y_{\ell+k}| y_{k+1, \ldots, k+\ell-1})
\right\}
 \right] \times
 \left[  
 \prod_{j=1}^d f_j(y_j)
 \right]. 
\end{multline*}
Alternatively, the joint density of the random variables $Y_1,\dots,Y_d$ can be expressed via a C-vine decomposition as
\begin{multline*}
f(y_1, \ldots, y_d) = \left[  
\prod_{\ell=1}^{d-1}\prod_{k=1}^{d-\ell}c_{\ell,\ell+k; 1, \ldots, \ell-1}
\left\{
F_{\ell | 1, \ldots, \ell-1} (y_{\ell} | y_{1, \ldots, \ell-1}),  \right. \right. \\
\left. \left. F_{\ell+k | 1, \ldots, \ell-1} (y_{\ell+k}| y_{1, \ldots, \ell-1}) )\right\}
 \right] \times
 \left[  
 \prod_{j=1}^d f_j(y_j)
 \right].
\end{multline*}


\subsection{Conditional Copulas and Vines}\label{sec:cop_vines}

Let us consider again $Y_1, \ldots, Y_d$, which are continuous random variables of interest and let $\textbf{X}=(X_1, \ldots, X_p)$ be a vector of covariates that may affect the dependence between $Y_1, \ldots, Y_d$.
Then, the conditional joint distribution function of $(Y_1, \ldots, Y_d)$ given $\textbf{X}= \textbf{x}$ is 
$$
F_x (y_1, \ldots, y_d) = P(Y_1 \leq y_1, \ldots, Y_d \leq y_d | \textbf{X}= \textbf{x}),
$$
under the assumption that such conditional distribution exists
(see \cite{gijbels2012multivariate}, \cite{abegaz2012semiparametric} and \cite{acar2011dependence}).

We denote the conditional marginals of $F_x$ as
\begin{eqnarray*}
F_{1,x}(y_1) & = & P(Y_1 \leq y_1 | \textbf{X}= \textbf{x}), \\
& & \ldots \\
F_{d,x}(y_d) & = & P(Y_d \leq y_d | \textbf{X}= \textbf{x}). 
\end{eqnarray*}
If the marginals are continuous, then Sklar's theorem allows us to write
$$
C_x (u_1, \ldots, u_d) = F_x \left( F_{1,x}^{-1} (u_1), \ldots, F_{d,x}^{-1} (u_d) \right)
$$
where $F_{j,x}^{-1} (u_j) = \inf \left\{ y_j : F_{j,x} \geq u_j \right\}$, for $j = 1,\ldots, d$, are the conditional quantile functions and $u_j = F_{j,x}(y_j)$ are called pseudo-observations or u-data.
The conditional copula $C_x$ fully describes the conditional dependence structure of $(Y_1, \ldots, Y_d)$ given $\textbf{X}= \textbf{x}$. 
Therefore, the conditional joint distribution of $(Y_1, \ldots, Y_d)$ given $\textbf{X}= \textbf{x}$ can be written as
$$
F_x (Y_1, \ldots, Y_d) = C_x \left( F_{1,x}(y_1), \ldots, F_{d,x} (y_d) \right).
$$

Let us denote the copula density corresponding to the distribution \\ $C_x \left( F_{1x}(y_1), \ldots, F_{dx} (y_d) \right)$ as
$$
c_x \left(u_1, \ldots, u_d \right) = c_{\boldsymbol{\theta}} (u_1, \ldots, u_d | \textbf{x}) = c_{\boldsymbol{\theta}(\textbf{x})} (u_1, \ldots, u_d),
$$
where $\boldsymbol{\theta}$ is the parameter vector of the $d$-variate copula density.
We assume that the function $\boldsymbol{\theta}(\textbf{x})$ depends on a vector of parameters $\boldsymbol{\beta}$ such that
\begin{equation}\label{eq:condcop}
c_{\boldsymbol{\theta}(\textbf{x})} (u_1, \ldots, u_d) = c_{\boldsymbol{\theta}(\textbf{x}| \boldsymbol{\beta})} (u_1, \ldots, u_d) = c_{1:d} (u_1, \ldots, u_d \,  |  \, \boldsymbol{\theta}(\textbf{x}| \boldsymbol{\beta})).
\end{equation}
The \eqref{eq:condcop} can be written in terms of vines, where each pair-copula depends on the vector of covariates $\textbf{X}$.
For example, considering the tri-variate vine copula introduced in Section \ref{sec:vines} we obtain
\begin{equation*}
\begin{split}
c_{1:3} (u_1,u_2, u_3 \,  |  \,  \boldsymbol{\theta}(\textbf{x}| \boldsymbol{\beta})) & =  c_{1,2}(F_{1,x}(y_1),F_{2,x}(y_2) \, | \, \boldsymbol{\theta}_{12}(\textbf{x}| \boldsymbol{\beta})) \\
& \times  c_{2,3}(F_{2,x}(y_2),F_{3,x}(y_3) \, | \, \boldsymbol{\theta}_{23}(\textbf{x}| \boldsymbol{\beta})) \\
& \times c_{1,3;2}(F_{1|2,x}(y_1|y_2),F_{3|2,x}(y_3|y_2) \,  |  \,  \boldsymbol{\theta}_{13;2}(\textbf{x}| \boldsymbol{\beta})) ,
\end{split}
\end{equation*}
where $\boldsymbol{\theta}_{13;2}(\textbf{x}| \boldsymbol{\beta})$ denotes the parameter function of the pair copula $c_{1,3;2}(\cdot,\cdot)$, $\boldsymbol{\theta}_{12}(\textbf{x}| \boldsymbol{\beta})$ denotes the parameter function of the pair copula $c_{1,2}(\cdot,\cdot)$ and $\boldsymbol{\theta}_{23}(\textbf{x}| \boldsymbol{\beta})$ denotes the parameter function of the pair copula $c_{2,3}(\cdot,\cdot)$.

In higher dimension, the conditional D-vine decomposition takes the form
\begin{multline*}
c_{1:d} (u_1, \ldots, u_d \,  |  \,  \boldsymbol{\theta}(\textbf{x}| \boldsymbol{\beta})) = \\
 \prod_{\ell=1}^{d-1}\prod_{k=1}^{d-\ell}c_{k,\ell+k; k+1, \ldots, k+\ell-1}
\left\{
F_{k | k+1, \ldots, k+\ell-1,x} (y_k | y_{k+1, \ldots, k+\ell-1}),  \right. \\
\left. F_{\ell+k | k+1, \ldots, k+\ell-1,x} (y_{\ell+k}| y_{k+1, \ldots, k+\ell-1}) \,  |  \,  \boldsymbol{\theta}_{k,\ell+k; k+1, \ldots, k+\ell-1}(\textbf{x}|
 \boldsymbol{\beta})\right\},
\end{multline*}
Moreover, the $d$-dimensional conditional C-vine is expressed as
\begin{multline*}
c_{1:d} (u_1, \ldots, u_d \,  |  \,  \boldsymbol{\theta}(\textbf{x}| \boldsymbol{\beta})) = \\
\prod_{\ell=1}^{d-1}\prod_{k=1}^{d-\ell}c_{\ell,\ell+k; 1, \ldots, \ell-1}
\left\{
F_{\ell | 1, \ldots, \ell-1,x} (y_{\ell} | y_{1, \ldots, \ell-1}),  \right. \\
\left. F_{\ell+k | 1, \ldots, \ell-1,x} (y_{\ell+k}| y_{1, \ldots, \ell-1}) \,  |  \,  \boldsymbol{\theta}_{\ell,\ell+k; 1, \ldots, \ell-1}(\textbf{x}|
 \boldsymbol{\beta})\right\},
\end{multline*}

\section{Dirichlet Process Mixture of Conditional Vine Copulas}



Vine copulas require a model selection step where a copula family is selected for each bivariate pair-copula forming the vine. However, the complexity of the problem increases with the vine dimension.
As a solution, we propose a hierarchical approach based on the vine construction.
More specifically, we adopt a Bayesian nonparametric approach, which overcomes the need of specifying the families of each pair-copula. Moreover, since often it is hard to consider the effect of the covariates on the dependence structures, particularly when the copula matrices are unstructured, we provide a general methodology which also allows to account for covariates in multivariate dependence structures, exploiting the flexibility of the pair-copula construction. More precisesely, we consider the effect of covariates on the depence structures between the paired variable, leveraging the information deriving from the covariates for a more accurate clustering.

Given a set of $N$ observations, let us consider the random vectors of interest $\textbf{Y}_1,\dots, \textbf{Y}_d$, each of dimension $N$, and the $N \times p$ matrix of observed covariates $\mathbb{X} = (\textbf{X}_1,\dots, \textbf{X}_p)$.
As in \cite{muller1997bayesian} and \cite{muller2010random}, we define the covariates as random variables, such that $\mathbb{X}$ corresponds to $N$ realizations of the independent random variables with densities $f_{h}(x_h)$, with $h=1,\dots,p$.

Starting from a simple example, let us consider the three-dimensional case with $d=3$ where $\textbf{Y}_1,\textbf{Y}_2,\textbf{Y}_3$ are random vectors of interest
and $(\textbf{X}_1,\dots, \textbf{X}_p)$ are covariate vectors influencing the dependence between the variables of interest.
Let $F_{1,x}(y_{1i}),F_{2,x}(y_{2i}),F_{3,x}(y_{3i})$, with $i=1,\dots,N$, be the conditional cdfs of the variables of interest,
 and let $\mathbb{X} = (\textbf{X}_1,\dots, \textbf{X}_p)$ be
  the $N \times p$ matrix of observed covariates, such that $x_{i,h}$ corresponds to the $i$-th realization of the $h$-th covariate. 
Adopting the $3$-dimensional D-vine specification illustrated in Figure \ref{fig:simp_vine},  we model the dependence between the variables of interest as the product of $d=3$ pair-copulas indexed by the vector of parameters $\boldsymbol{\beta}=\left ( \beta_{0_{12}} ,\beta_{1_{12}},\dots,\beta_{p_{12}},\beta_{0_{23}},\beta_{1_{23}},\dots,\beta_{p_{23}},\beta_{0_{13;2}},\beta_{1_{13;2}},\dots,\beta_{p_{13;2}} \right)$, where the subscripts $\{12\}, \{23\},$ $\{13;2\}$ denote the pair-copulas in the vine. 
We assume that the covariate distributions are governed by the matrix of parameters $\boldsymbol{\phi}=\left ( \phi_1,\dots, \phi_p \right)$ where the dimension of $\boldsymbol{\phi}$ depends on the covariate kernels. 
For example, if we assume each covariate $X_h$ to be Normally distributed with parameters $(\mu_h,\sigma_h)$, for $h=1,\dots,p$, $\boldsymbol{\phi}$ will be a $(p\times2)$-dimensional matrix. The vector of parameters $\boldsymbol{\xi} = \left( \boldsymbol{\beta}, \boldsymbol{\phi}\right)$ is defined on the parameter space $\Xi$. 
Let $G$ be a probability distribution defined on $\Xi$. Let us also assume that $\textbf{f}(\textbf{x})=\prod_{h=1}^{p}f_{h}(x_h)$ denotes the product of the densities of the covariates, and $c_{\boldsymbol{\xi}}(\cdot,\cdot,\cdot |\textbf{x})$ denotes the $3$-variate conditional copula density arising from the vine. We rewrite the density $\textbf{f}_{G}(\textbf{x})\cdot c_G(\cdot,\cdot,\cdot |\textbf{x})$ as an infinite mixture of vine copulas with kernel $\textbf{f}_{\boldsymbol{\xi}}(\textbf{x})\cdot c_{\boldsymbol{\xi}}(\cdot,\cdot,\cdot |\textbf{x})$ with respect to the mixing measure $G$, such that
\begin{multline*}
\textbf{f}_{G}(\textbf{x})c_{G}(F_{1,x}(y_1),F_{2,x}(y_2)|\textbf{x}) c_{G}(F_{2,x}(y_2),F_{3,x}(y_3)|\textbf{x}) \\
\times  c_{G}(F_{1|2,x}(y_1|y_2),F_{3|2,x}(y_3|y_2)|\textbf{x})= \\
  \int \textbf{f}_{\boldsymbol{\xi}}(\textbf{x}) c_{\boldsymbol{\xi}}(F_{1,x}(y_1),F_{2,x}(y_2)| \textbf{x})  c_{\boldsymbol{\xi}}(F_{2,x}(y_2),F_{3,x}(y_3) | \textbf{x}) \\
\times   c_{\boldsymbol{\xi}}(F_{1|2,x}(y_1|y_2),F_{3|2,x}(y_3|y_2) |  \textbf{x}) dG(\boldsymbol{\xi}).
\end{multline*}

Generalizing, let us consider $d$-variables of interest $Y_1,\dots,Y_d$. 
Given a set of $N$ observations, let us assume that $F_{1,x}(y_{1i}),\dots,F_{d,x}(y_{di})$, with $i=1,\dots,N$, are the conditional cdfs of the $d$ variables of interest.
The multivariate dependence structure is specified by a vine defined as the product of $\nu=d(d-1)/2$ pair copulas, indexed by the $\nu\times(q+1)$-dimensional vector of parameters $\boldsymbol{\beta}$,
while $f_{h}(x_h)$, $h=1,\dots,p$, are independent random variables with parameters $\boldsymbol{\phi}=\left (\phi_{1},\dots,\phi_p \right)$. Note that $q\ge p$ and its value depends on the chosen link function; for example if the link is linear $q=p$. Let the vector of parameters $\boldsymbol{\xi}=\left (\boldsymbol{\beta},\boldsymbol{\phi} \right)$ be defined on the parameter space $\Xi$. Again, let $G$ be a probability distribution defined on $\Xi$. 
As before, we rewrite the density $\textbf{f}_{G}(\textbf{x}) \cdot c_G(\cdot, \ldots, \cdot | \textbf{x})$ as an infinite mixture of conditional vine copulas with kernel $\textbf{f}_{\boldsymbol{\xi}}(\textbf{x})\cdot c_{\boldsymbol{\xi}}(\cdot, \ldots, \cdot | \textbf{x})$ with respect to the mixing measure $G$, that in the $D$-vine case is
\begin{multline*}
\textbf{f}_{G}(\textbf{x})\prod_{\ell=1}^{d-1}\prod_{k=1}^{d-\ell}c_{G}\left (F_{k | k+1, \ldots, k+\ell-1,x} (y_k | y_{k+1, \ldots, k+\ell-1}),  \right. \\
\left. F_{\ell+k | k+1, \ldots, k+\ell-1,x} (y_{\ell+k}| y_{k+1, \ldots, k+\ell-1}) | \textbf{x}\right ) = \\
\int \textbf{f}_{\boldsymbol{\xi}}(\textbf{x})\prod_{\ell=1}^{d-1}\prod_{k=1}^{d-\ell}c_{\boldsymbol{\xi}}\left (F_{k | k+1, \ldots, k+\ell-1,x} (y_k | y_{k+1, \ldots, k+\ell-1}),   \right. \\
\left. F_{\ell+k | k+1, \ldots, k+\ell-1,x} (y_{\ell+k}| y_{k+1, \ldots, k+\ell-1}) | \textbf{x}\right ) dG(\boldsymbol{\xi}) = \\
\int \textbf{f}_{\boldsymbol{\phi}}(\textbf{x})\prod_{\ell=1}^{d-1}\prod_{k=1}^{d-\ell}c_{\boldsymbol{\beta}}\left (F_{k | k+1, \ldots, k+\ell-1,x} (y_k | y_{k+1, \ldots, k+\ell-1}),  \right. \\
\left. F_{\ell+k | k+1, \ldots, k+\ell-1,x} (y_{\ell+k}| y_{k+1, \ldots, k+\ell-1}) | \textbf{x}\right ) dG(\boldsymbol{\phi,\beta}).
\end{multline*}
In the $C$-vine case we have
\begin{equation*}
\begin{gathered}
\textbf{f}_{G}(\textbf{x})\prod_{\ell=1}^{d-1}\prod_{k=1}^{d-\ell}c_{G}\left (F_{\ell | 1, \ldots, \ell-1,x}(y_{\ell} | y_{1, \ldots, \ell-1}),F_{\ell+k | 1, \ldots, \ell-1,x} (y_{\ell+k}| y_{1, \ldots, \ell-1}) |\textbf{x}\right ) = \\
\int \textbf{f}_{\boldsymbol{\xi}}(\textbf{x})\prod_{\ell=1}^{d-1}\prod_{k=1}^{d-\ell}c_{\boldsymbol{\xi}}\left (F_{\ell | 1, \ldots, \ell-1,x}(y_{\ell} | y_{1, \ldots, \ell-1}),F_{\ell+k | 1, \ldots, \ell-1,x} (y_{\ell+k}| y_{1, \ldots, \ell-1}) |\textbf{x}\right ) dG(\boldsymbol{\xi}) = \\
\int \textbf{f}_{\boldsymbol{\phi}}(\textbf{x})\prod_{\ell=1}^{d-1}\prod_{k=1}^{d-\ell}c_{\boldsymbol{\beta}}\left (F_{\ell | 1, \ldots, \ell-1,x}(y_{\ell} | y_{1, \ldots, \ell-1}),F_{\ell+k | 1, \ldots, \ell-1,x} (y_{\ell+k}| y_{1, \ldots, \ell-1}) |\textbf{x}\right ) dG(\boldsymbol{\phi,\beta}).
\end{gathered}
\end{equation*}

With a Dirichlet Process (DP) prior on the mixing measure $G$, we get a Dirichlet Process Mixture (DPM) of conditional vine copulas, wich may be rewritten in hierarchical form as follows. Let $\textbf{y}_i=(y_{1i},\dots,y_{di})$ denote the vector containing the $i$-th elements of each variable $\textbf{Y}_j$ for $j=1,\dots,d$, and let $\textbf{y}_{-i}=(y_{1,-i},\dots,y_{d,-i})$ denote the $(N-1)\times d$ elements obtained excluding the $i$-th observation of each $\textbf{Y}_j$ for $j=1,\dots,d$. 
Further, let $(\textbf{u}_i|\textbf{u}_{-i}) = F_{i,x}(\textbf{y}_i|\textbf{y}_{-i})$ denote the vector 
of pseudo-observations 
for each $i=1,\dots,N$. 
Also, let $\textbf{x}_i=(x_{1i},\dots,x_{pi})$ be the vector containing the $i$-th elements of each covariate $\textbf{X}_h$ for $h=1,\dots,p$.
Then we define a DPM model of conditional vine copulas as
\begin{equation*}
\begin{aligned}
 & (\textbf{u}_i|\textbf{u}_{-i}) | (\boldsymbol{\beta}_i,\textbf{x}_i)  \stackrel{ind}{\sim} c_{\boldsymbol{\beta}_i}(\cdot, \ldots, \cdot | \textbf{x}_i) \quad i=1,...,N, \\
 &  \textbf{f}({\textbf{x}_{i}})|\boldsymbol{\phi}_i  \stackrel{ind}{\sim}\textbf{f}_{\boldsymbol{\phi}_i},\\
 & (\boldsymbol{\phi}_i,\boldsymbol{\beta}_i)|G  \stackrel{iid}{\sim} G,\\
 & G  \sim DP(M,G_0).
\end{aligned}
\end{equation*}
with total mass parameter $M$ and centering measure $G_0$. 
The posterior distribution $\Pi(G| \textbf{Y},\textbf{X})$ is a mixture of DP models, mixing with respect to the latent variables $\boldsymbol{\xi}_i$ specific to each observation $i$ 
for $i=1,\dots,N$:
\begin{equation*}
G| \textbf{Y},\textbf{X} \sim\int DP\left (MG_0+\sum_{i=1}^N\delta_{\boldsymbol{\phi}_i\boldsymbol{\beta}_i } \right )d\Pi(\boldsymbol{\phi},\boldsymbol{\beta}|\textbf{y},\textbf{x}),
\end{equation*}
where $\delta_t$ denotes the Dirac measure at $t$.

In general, the choice of the kernel for DPM models should consider two aspects. From a computational point of view, the conjugacy between the centering measure $G_0$ and the kernel $f_{\boldsymbol{\phi}} c_{\boldsymbol{\beta}}$ is particularly convenient. A more important feature is the flexibility of the chosen density. In our case, we need to specify a product of bivariate copula densities and a product of independent densities for the covariates. The choice of the covariate densities will depend on the nature of $X_h$, $h=1,\dots,p$. On the other hand, in order to model the dependence structure between the variables, we need a copula which is able to capture various kinds of dependence and may approximate different copula families. 
\cite{wu2015bayesian} showed that bivariate density functions on the real plain can be arbitrarily well approximated by a mixture of a countably infinite number of bivariate normal distributions. \cite{valle2018bayesian} proposed a Bayesian nonparametric estimation of bivariate conditional copulas, with a Gaussian copula as kernel of a DPM. 
Following the previous approaches, we propose as kernel the product of the density of a Gaussian vine-copula and the densities of covariates which depend on the nature of $\mathbb{X}$.

\section{MCMC sampling for DPM of Conditional Vine Copulas}
One of the main advantages of using nonparametric methods is the ability to reduce uncertainty avoiding distributional assumptions. However, in the Bayesian framework the increased flexibility has a computational cost. 
In DPM models there are two levels of conjugacy: between the DP random measure $\Pi(G)$ and its posterior $\Pi(G|\mathbf{y})$, and between the kernel element $\textbf{f}_{\boldsymbol{\xi}}c_{\boldsymbol{\xi}}$ and the centering measure $G_0$, which plays the role of prior on $\boldsymbol{\xi}$. When considering non-conjugate DP mixtures, we refer to the case where $c_{\boldsymbol{\xi}}$ and $G$ are not conjugate; a sampler for this case is the \textit{no gaps} sampler, proposed by \cite{maceachern1998estimating}, which still relies on conjugacy of the DP posterior. 
In our case, we choose as kernel the product of the density of a Gaussian vine-copula and the densities of the covariates which depend on the nature of $\mathbb{X}$. However, we lose the conjugacy between the kernel and the centering measure $G_0$ because of 
the presence of the covariates.

Let $(u_{1i},\dots,u_{di}) = (F_{1,x}(y_{1i}),\dots,F_{d,x}(y_{di}))$, for $i=1,\dots,N$, 
be pseudo-observations defined in the hypercube $\mathbf{I}^d$. Let $\mathbb{X}$ be the $N \times p$ matrix of covariates. The Gaussian bivariate copula is governed by the correlation parameter $\rho\in(-1,1)$. 
Following the conditional copula literature,
the correlation parameter is associated to
the covariates through a link function $g$, such that  
$$\rho (\textbf{x} | \boldsymbol{\beta}) = g^{-1} ( \eta(\textbf{x} | \boldsymbol{\beta}) )$$
where $g^{-1}$ is the inverse link function (that we assume to be the Fisher's transform) and $\eta(\cdot)$ is a calibration function.
In our setting, we need to define a centering measure for the parameters of the $\nu=d(d-1)/2$ pair-copulas forming the vine and the parameters of the covariate densities. The centering measure dimension will depend on the number of covariates and the probability models assumed for the covariates. Hence, we need to define a $(\nu \times q+r)$-dimensional centering measure, where $q$ represents the number of unknown parameters of the calibration function for each pair copula and $r$ denotes the total number of unknown parameters of the covariate densities.

Considering the general $D$-vine construction, 
we assume, as kernel density of the mixture, the product of the bivariate densities
\begin{multline*}
\textbf{f}_{\boldsymbol{\phi}}(\textbf{x})\prod_{\ell=1}^{d-1}\prod_{k=1}^{d-\ell}c_{\boldsymbol{\beta}}\left (F_{k | k+1, \ldots, k+\ell-1,x} (y_k | y_{k+1, \ldots, k+\ell-1}),  \right. \\
\left. F_{\ell+k | k+1, \ldots, k+\ell-1,x} (y_{\ell+k}| y_{k+1, \ldots, k+\ell-1}) | \textbf{x}\right ).
\end{multline*}
In the $C$-vine case, instead, the kernel density is
\begin{equation*}
\textbf{f}_{\boldsymbol{\phi}}(\textbf{x})\prod_{\ell=1}^{d-1}\prod_{k=1}^{d-\ell}c_{\boldsymbol{\beta}}\left (F_{\ell | 1, \ldots, \ell-1,x}(y_{\ell} | y_{1, \ldots, \ell-1}),F_{\ell+k | 1, \ldots, \ell-1,x} (y_{\ell+k}| y_{1, \ldots, \ell-1}) |\textbf{x}\right ).
\end{equation*}

In both cases, the model parameters are included in the $(\nu \times q + r)$-dimensional random vector $\boldsymbol{\xi}=\left (\boldsymbol{\beta},\boldsymbol{\phi} \right )$. Since the discreteness of the DP implies a positive probability for ties among the latent elements of the vector $\boldsymbol{\xi}=\left (\boldsymbol{\beta},\boldsymbol{\phi} \right )$, the DPM induces a probability model on clusters. Formally, let $\xi_m^*=\left (\beta_m^*,\phi_m^* \right )$, $m=1,\dots,n$, denote $n \le N$ unique values of the vector, $\Psi_m=\left\{ \iota : \xi_\iota=\xi_m^*\right\}$, and let $N_m=|\Psi_m|$ denote the number of $\xi_\iota$ tied with $\xi_m^*$. Since the $\xi_\iota$ are random, also the $\Psi_m$ are random. Thus, the DPM implies a model on the random partition $\zeta_i=\{ \Psi_1,\dots,\Psi_n\}$ of the experimental units $\{1,\dots,N\}$; the posterior model $\pi(\boldsymbol{\psi}|\textbf{Y},\textbf{X})$ reports posterior inference on clustering of the data. 
We represent the clustering by an equivalent set of cluster membership indicators, $\psi_i=m$ if $i \in \Psi_m$, with clusters labeled by order of appearance. 

Let $\xi_m^{*-}=\left (\beta_m^{*-},\phi_m^{*-} \right )$ denote the $m$-th of the $n^-$ unique values among $\boldsymbol{\xi}_{-i}=\left (\boldsymbol{\beta}_{-i},\boldsymbol{\phi}_{-i} \right )$, which represents the vector $\boldsymbol{\xi}=\left (\boldsymbol{\beta},\boldsymbol{\phi} \right )$ of the $n \le N$ unique values without the $i$-th element $\xi_i=\left(\beta_i,\psi_i \right)$. Let $\Psi_m=(\iota : {\xi}_\iota={\xi}_m^*)$, so that, if $\psi_i=m$, the $i$-th observation belongs to the $m$-th cluster and $\omega_m$ represents the number of observations lying inside the cluster $m$.
We will now describe the main steps of the implementation of the algorithm.
\vspace{0.5cm}

1. We first define $\pi(\psi_i=m|\boldsymbol{\psi}_{-i},\xi_m^*,\textbf{X},\textbf{Y})$ for $m=1,\dots, n^-$ and $m=n^-+1$. As proved by \cite{maceachern1998estimating}, the probability that the $i$-th element belongs to the $m$-th cluster is
\begin{equation*}
\begin{cases}
N_m^{-}\textbf{f}_{\phi^*_m}(\textbf{x})c_{\beta^*_m}(\textbf{u} | \textbf{x}) \quad \text{for } m=1,\dots,n^{-} \\
\frac{M}{n^- + 1}\textbf{f}_{\phi_{n^- +1,1}^*}(\textbf{x}) c_{\beta_{n^- +1,1}^*}(\textbf{u} | \textbf{x})\quad \text{for }  m=n^-+1
\end{cases},
\end{equation*}
where $N_m^-$ is the number of elements in the $m$-th cluster with exclusion of the $i$-th observation, $M$ represents the precision parameter of the DP, $\textbf{f}_{\phi}$ is the $p$-dimensional product of densities of the covariates and $c_{\beta}$ is the $\nu$-dimensional product of Gaussian bivariate conditional copula densities. \\

\vspace{0.5cm}

2. We compute the conditional distribution functions $F_{a|b,x}(y_a|y_b)$, following \cite{joe1996families} as
\begin{equation}
F_{a|b,x}(y_a|y_b)=\frac{\partial C_{\beta_m^*}(F_{a,x}(y_{a_m}^*),F_{b,x}(y_{b_m}^*)|\textbf{x})}{\partial F_{b,x}(y_{b_m}^*)}\quad\text{for }m=1,\dots,n,
\end{equation}
for $b=1\dots d-1$ and for $a=1\dots d-b$. We consider the belonging cluster $m$ in the derivatives for each variable.\\

\vspace{0.5cm}

3. Finally, for each cluster $m$ the likelihood function is
\begin{equation*}
\mathcal{L}(\beta_{m}^*;\phi_{m}^*;\Psi_m)=\prod_{\kappa=1}^{N_m} \textbf{f}_{\phi_m^{*}}(\textbf{x}_{m_{\kappa}}|\Psi_m)c_{\beta_m^*}(F(\textbf{y}_{m_{\kappa}})|\textbf{x},\Psi_m),
\end{equation*}
where $c_{\beta}(\cdot)$ represents the vine copula density.
We compute the posterior density for $\xi_m^*$ as
\begin{equation*}
\pi(\xi_{m}^*|\boldsymbol{\psi},\textbf{y},\textbf{x})=G_0(\xi_{m}^*)\cdot\mathcal{L}(\xi_{m}^*;\Psi_m)
\end{equation*}
or equivalentely 
\begin{equation*}
\pi(\beta_{m}^*,\phi_{m}^*|\boldsymbol{\psi},\textbf{y},\textbf{x})= G_0(\beta_{m}^*,\phi_{m}^*)\mathcal{L}(\phi_{m}^*;\beta_{m}^*,\Psi_m),
\end{equation*}
which is a $(\nu \times q + r)$-dimensional density approximated with a Metropolis-Hastings step. The algorithm is summarized below.

\begin{algorithm}
\caption{\textbf{MCMC sampler for DPM of conditional vine copulas}}
\vspace{0.2cm}

$\bullet$ \textbf{Clustering}:  \\
$-$ for $i =1\dots,N$ draw $\psi_i\sim\pi(\psi_i=M|\boldsymbol{\psi}_{-i},\beta_M^*,\phi_M^*,\textbf{y},\textbf{x})$;\\
$\bullet$ \textbf{Cluster parameters}:  \\
$\circ$ update the covariates parameters: \\
$-$ for $m=1,\dots,n$ draw $\phi_{m}^*\sim\pi(\phi_{m}^*|\boldsymbol{\psi},\textbf{x})$\\
$-$ for $m=n+1,\dots,N$ draw $\phi_{m}^*\sim G_0$.\\
$\circ$ compute the conditioned distribution functions; \\
$\circ$ for each pair-copula: \\
$-$ for $m=1,\dots,n$ draw $\beta_{m}^*\sim\pi(\beta_{m}^*|\boldsymbol{\psi},\textbf{y},\textbf{x})$;\\
$-$ for $m=n+1,\dots,N$ draw $\beta_{m}^*\sim G_0$.
\end{algorithm}

\section{Application}

In this Section we apply the proposed model to both simulated and real data. In particular, we start with a thorough simulation study in order to check the accuracy of the methodology for clustering and density estimation. 

\subsection{Simulation Study}

We carried out five different simulation experiments drawing, for each one of them, 100 samples of size 100 from $3$-dimensional conditional vine copulas whose graphical structure is represented in Figure \ref{fig:simp_vine}.
For each sample we generated a Normally distributed covariate $X \sim \text{N}(\mu_x,\sigma_x)$. 
We defined the total mass parameter as $M=1$ and the centering measure as 
$$
G_0\equiv \text{N}\left (\mu_{0_x},\sigma^{2}_{0_x}\right )\times\text{Inv-Gamma}(a,b)\times \text{N}_{(3\times q)}(\boldsymbol{\mu},\boldsymbol{\Sigma}) 
$$
where, again, $q$ is the number of unknown parmeters of the calibration function.
In the simulation experiments, for each bivariate copula forming the vine, we considered  the following linear and non-linear calibration functions: 
 \begin{enumerate}
\item $\eta_{s}(x | \boldsymbol{\beta})=\beta_{0_{s}}+\beta_{1_{s}}x$,

\item $\eta_{s}(x| \boldsymbol{\beta})=\beta_{0_{s}}+\beta_{1_{s}}x+\beta_{2_{s}}e^{-\beta_{3_{s}}x}$,
\end{enumerate}
where the subscript $s$ denotes the pair-copulas $\{ 12 \}$, $\{ 23 \}$ and $\{ 13;2 \}$ in the vine.

In each simulation experiment we generated samples from the posterior distributions of the model parameters via Gibbs sampling with 5000 iterations and a burnin of 1000.

\begin{figure}[htbp]
  \centering
    \includegraphics[width=1\linewidth]{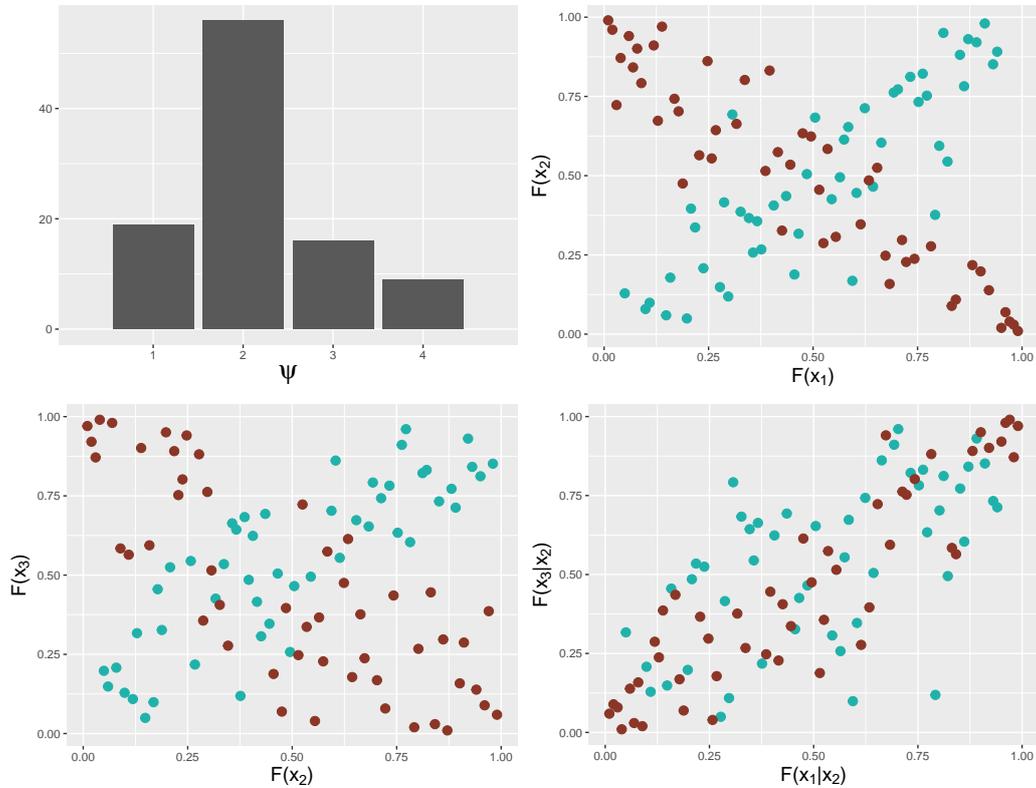}
     \caption{Results from the first scenario, with 100 samples drawn from a $3$-dimensional two-component mixture of conditional Gaussian vine copulas. 
     The covariate was generated from a N(1, 0.5) using the first calibration function with parameters $\boldsymbol{\beta}|(\psi=1)=(\beta_{0_{12}},\beta_{1_{12}},\beta_{0_{23}}, \beta_{1_{23}},\beta_{0_{13;2}}, \beta_{1_{13;2}})=(1,0.5,0.5,0.3,0.5,0.5)$ and $\boldsymbol{\beta}|(\psi=2)=(\beta_{0_{12}},\beta_{1_{12}},\beta_{0_{23}}, \beta_{1_{23}},\beta_{0_{13;2}}, \beta_{1_{13;2}})=(-1,-0.5,-0.5,-0.3,0.5,0.5)$. 
The top left panel shows the barplot of the mode of the number of observed mixture components.
The top right and the bottom panels show the scatterplots of the simulated u-data for one of the 100 samples; the blue points denote the observations simulated from the first component $\psi=1$, while the brown points denote the observations simulated from the second component $\psi=2$.}
 \label{Clustering1}
\end{figure}

In order to illustrate the clustering performance of the proposed methodology, in the first scenario we drew observations from a two-component mixture of conditional Gaussian vine copulas. The covariate was generated from a N(1, 0.5) adopting the first calibration function. 
We assumed the following parameters for the first component of the mixture
$$
\boldsymbol{\beta}|(\psi=1)=(\beta_{0_{12}},\beta_{1_{12}},\beta_{0_{23}}, \beta_{1_{23}},\beta_{0_{13;2}}, \beta_{1_{13;2}})=(1,0.5,0.5,0.3,0.5,0.5),
$$ 
while for the second component of the mixture we assumed
$$
\boldsymbol{\beta}|(\psi=2)=(\beta_{0_{12}},\beta_{1_{12}},\beta_{0_{23}}, \beta_{1_{23}},\beta_{0_{13;2}}, \beta_{1_{13;2}})=(-1,-0.5,-0.5,-0.3,0.5,0.5).
$$
The top left panel of Figure \ref{Clustering1} shows the barplot of the mode of the number of mixture components, which concentrates on the true value.
The top right and the bottom panels of Figure \ref{Clustering1} depict the scatterplots of the simulated u-data for one of the 100 samples. The blue points denote the observations simulated from the first component $\psi=1$, while the brown points denote the observations simulated from the second component $\psi=2$.
The results demonstrate that the proposed approach is appropriate for clustering purposes.


\begin{figure}[htbp]
  \centering
    \includegraphics[width=1\linewidth]{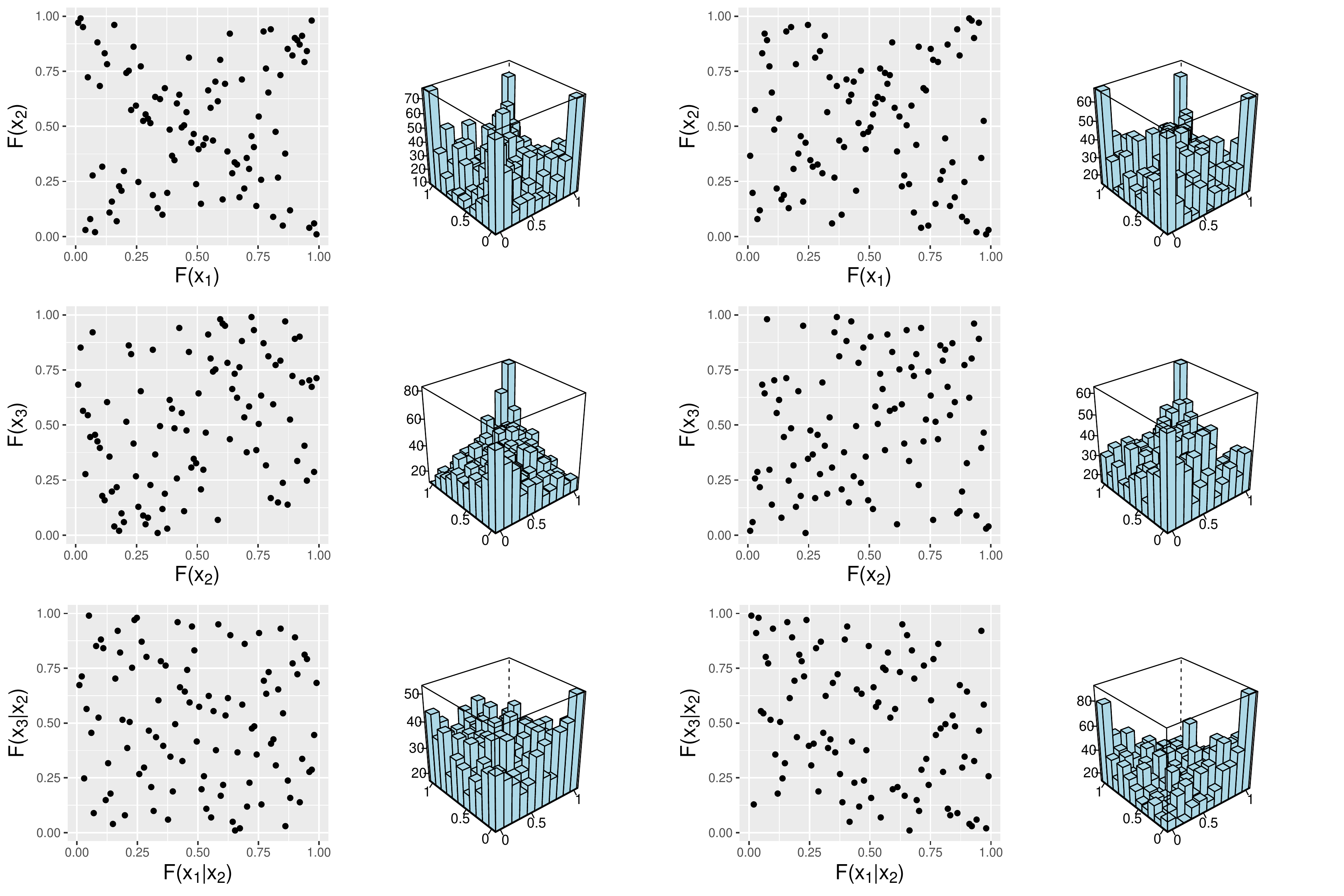}
      \caption{Results from the second scenario, where scatterplots show simulated u-data and 3-D histograms show the predicted u-data. The covariate was generated from a N(1, 1) using the first calibration function.
The left panels depict the results for a two equally weighted components mixture of conditional Gaussian vine copulas with parameters $\boldsymbol{\beta}|(\psi=1)=(\beta_{0_{12}},\beta_{1_{12}},\beta_{0_{23}}, \beta_{1_{23}},\beta_{0_{13;2}}, \beta_{1_{13;2}})=(0.4,0.7,-0.3,0.5,-0.1,-0.1)$ and $\boldsymbol{\beta}|(\psi=2)=(\beta_{0_{12}},\beta_{1_{12}},\beta_{0_{23}}, \beta_{1_{23}},\beta_{0_{13;2}}, \beta_{1_{13;2}})=(-0.4,-0.7,0.5,-0.3,0.1,0.1)$. 
The right panels depict the results of a two components mixture of conditional Gaussian vine copulas with equal weights and parameters $\boldsymbol{\beta}|(\psi=1)=(\beta_{0_{12}},\beta_{1_{12}},\beta_{0_{23}}, \beta_{1_{23}},\beta_{0_{13;2}}, \beta_{1_{13;2}})=(1,0.4,-0.8,0.5,-0.3,-0.2)$ and $\boldsymbol{\beta}|(\psi=2)=(\beta_{0_{12}},\beta_{1_{12}},\beta_{0_{23}}, \beta_{1_{23}},\beta_{0_{13;2}}, \beta_{1_{13;2}})=(-0.4,-0.3,0,0.5,0.1,0.1)$.}
 \label{DensEst1}
\end{figure}

\begin{figure}[htbp]
     \centering
\vspace{1cm}
    \includegraphics[width=1\linewidth]{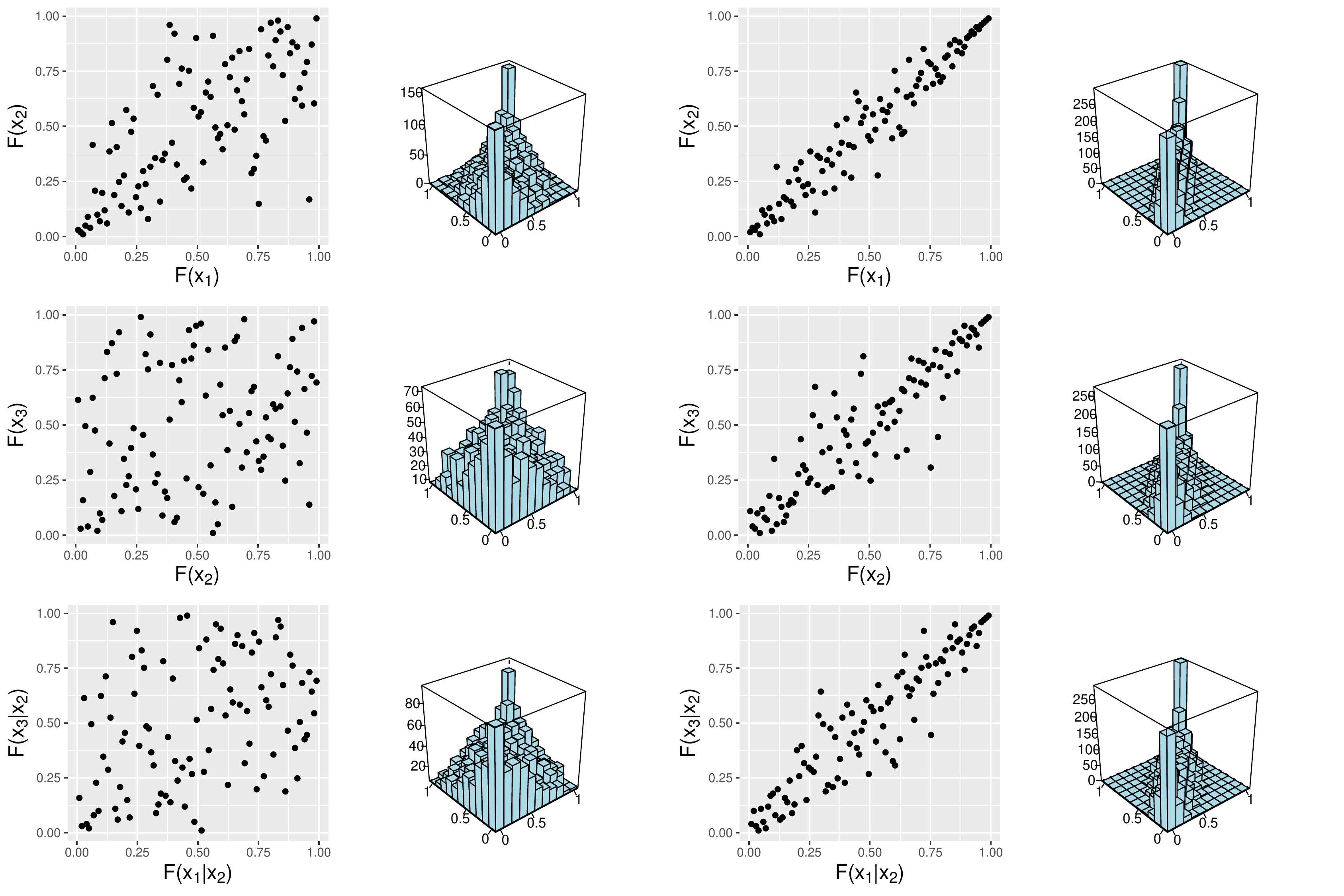}
       \caption{Results from the third scenario, where scatterplots show simulated u-data and 3-D histograms show the predicted u-data. The covariate was generated from a N(1, 1) using the first calibration function.
The left panels depict the results for a two equally weighted components mixture of conditional Clayton vine copulas with parameters $\boldsymbol{\beta}|(\psi=1)=(\beta_{0_{12}},\beta_{1_{12}},\beta_{0_{23}}, \beta_{1_{23}},\beta_{0_{13;2}}, \beta_{1_{13;2}})=(0.3,0.7,0,0.5,0.2,0)$ and $\boldsymbol{\beta}|(\psi=2)=(\beta_{0_{12}},\beta_{1_{12}},\beta_{0_{23}}, \beta_{1_{23}},\beta_{0_{13;2}}, \beta_{1_{13;2}})=(0.3,0.2,0,0.1,0.2,0)$. 
The right panels depict the results of a two-component mixture of conditional Gumbel vine copulas with equal weights and parameters $\boldsymbol{\beta}|(\psi=1)=(\beta_{0_{12}},\beta_{1_{12}},\beta_{0_{23}}, \beta_{1_{23}},\beta_{0_{13;2}}, \beta_{1_{13;2}})=(1.3,0.7,1.0,0.5,1.2,0)$ and $\boldsymbol{\beta}|(\psi=2)=(\beta_{0_{12}},\beta_{1_{12}},\beta_{0_{23}}, \beta_{1_{23}},\beta_{0_{13;2}}, \beta_{1_{13;2}})=(1.3,0.2,1.0,0.1,1.2,0)$.}
   \label{DensEst2}
\end{figure}

\begin{figure}[htbp]
\centering
     \centering
 \includegraphics[width=1\linewidth]{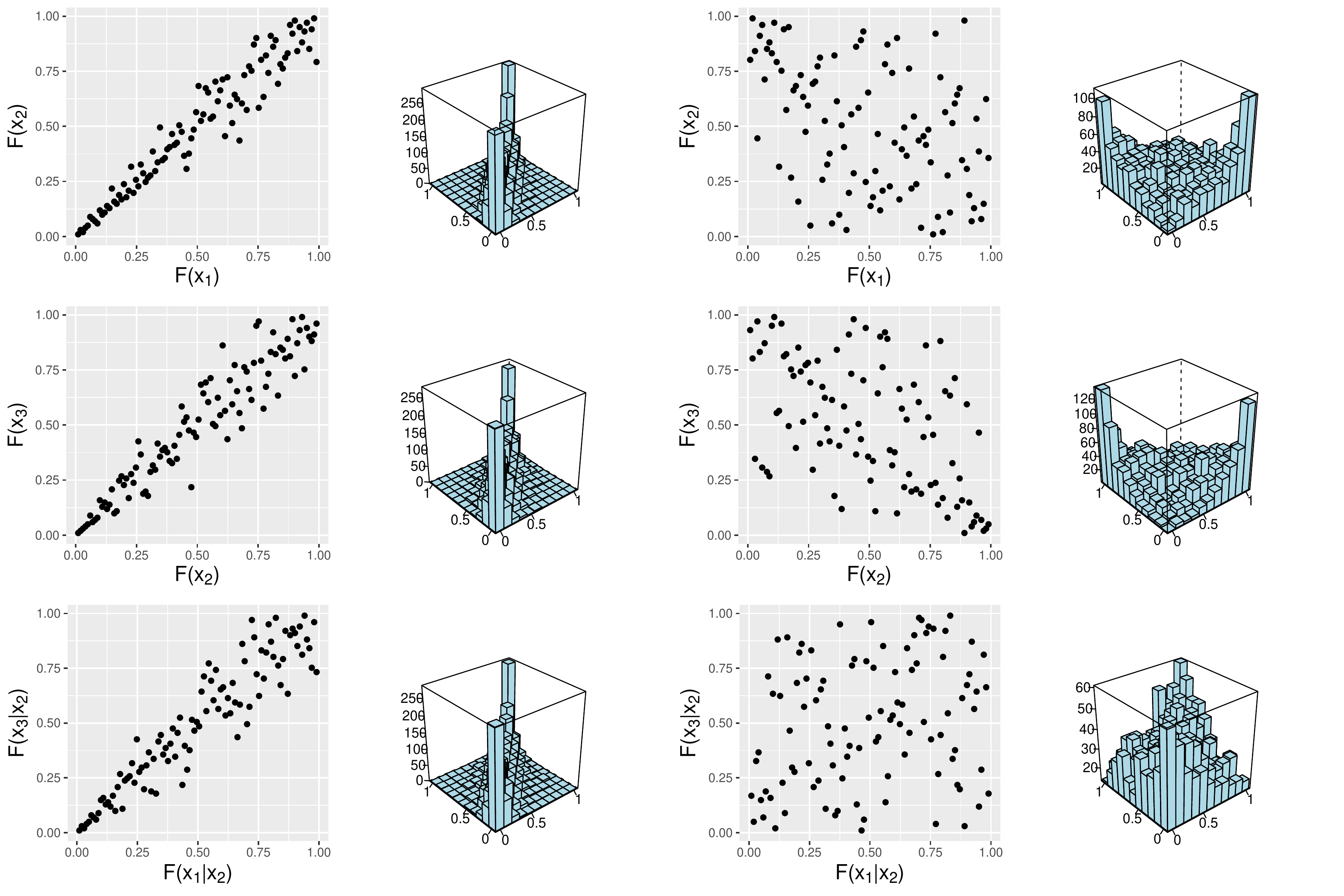}
  \caption{Results from the fourth scenario, where scatterplots show simulated u-data and 3-D histograms show the predicted u-data. The covariate was generated from a N(1, 1) using the first calibration function.
The left panels depict the results for a conditional Clayton vine copula with parameters $\boldsymbol{\beta}=(\beta_{0_{12}},\beta_{1_{12}},\beta_{0_{23}}, \beta_{1_{23}},\beta_{0_{13;2}}, \beta_{1_{13;2}})=(1,0.2,0.8,0.3,0.4,0.1)$.
The right panels depict the results for a conditional vine with rotated Clayton copulas in the first tree and a Clayton copula in the second tree, with parameters 
$\boldsymbol{\beta}=(\beta_{0_{12}},\beta_{1_{12}},\beta_{0_{23}}, \beta_{1_{23}},\beta_{0_{13;2}}, \beta_{1_{13;2}})=(-1.5,-0.4,-0.9,-0.8, 0.4,0.6)$.}
\label{DensEst3}
\end{figure}

In order to demonstrate the flexibility of our approach for density estimation, we compared simulated and predicted samples. 
More precisely, in the second, third and fourth scenarios, we generated data from conditional vine copulas constructed with various copula families and different parameters. In these three scenarios, we adopted the first calibration function, simulating the covariate from a N(1, 1).
The results are presented in Figures \ref{DensEst1}, \ref{DensEst2} and \ref{DensEst3}, where the scatterplots show the simulated u-data samples, while the 3-D histograms show the predicted u-data samples.

Figure \ref{DensEst1} illustrates the results obtained for scenario 2, where we generated observations from two different two-components mixtures of conditional Gaussian vine copulas. 
The left panels show the results for a two equally weighted components mixture. 
The parameters of the calibration function for the first mixture component are 
$$
\boldsymbol{\beta}|(\psi=1)=(\beta_{0_{12}},\beta_{1_{12}},\beta_{0_{23}}, \beta_{1_{23}},\beta_{0_{13;2}}, \beta_{1_{13;2}})=(0.4,0.7,-0.3,0.5,-0.1,-0.1)
$$
while the parameters for the second component are
$$
\boldsymbol{\beta}|(\psi=2)=(\beta_{0_{12}},\beta_{1_{12}},\beta_{0_{23}}, \beta_{1_{23}},\beta_{0_{13;2}}, \beta_{1_{13;2}})=(-0.4,-0.7,0.5,-0.3,0.1,0.1).
$$
The right panels show the results for another two components mixture with equal weights and parameters 
$$
\boldsymbol{\beta}|(\psi=1)=(\beta_{0_{12}},\beta_{1_{12}},\beta_{0_{23}}, \beta_{1_{23}},\beta_{0_{13;2}}, \beta_{1_{13;2}})=(1,0.4,-0.8,0.5,-0.3,-0.2)
$$ 
for the first component, and parameters
$$
\boldsymbol{\beta}|(\psi=2)=(\beta_{0_{12}},\beta_{1_{12}},\beta_{0_{23}}, \beta_{1_{23}},\beta_{0_{13;2}}, \beta_{1_{13;2}})=(-0.4,-0.3,0,0.5,0.1,0.1)
$$
for the second component.
The results show a good performance in terms of density estimation for the proposed methodology in the Gaussian case.

Figure \ref{DensEst2} shows the results obtained for scenario 3, where observations were generated from a  two equally weighted components mixture of conditional Clayton vine copulas (left panels), and from a two equally weighted components mixture of conditional Gumbel vine copulas (right panel).
The parameters of the conditional Clayton vine copula mixture are
$$
\boldsymbol{\beta}|(\psi=1)=(\beta_{0_{12}},\beta_{1_{12}},\beta_{0_{23}}, \beta_{1_{23}},\beta_{0_{13;2}}, \beta_{1_{13;2}})=(0.3,0.7,0,0.5,0.2,0),
$$
for the first component and 
$$
\boldsymbol{\beta}|(\psi=2)=(\beta_{0_{12}},\beta_{1_{12}},\beta_{0_{23}}, \beta_{1_{23}},\beta_{0_{13;2}}, \beta_{1_{13;2}})=(0.3,0.2,0,0.1,0.2,0),
$$
for the second component.
The parameters of the conditional Gumbel vine copula mixture are
$$
\boldsymbol{\beta}|(\psi=1)=(\beta_{0_{12}},\beta_{1_{12}},\beta_{0_{23}}, \beta_{1_{23}},\beta_{0_{13;2}}, \beta_{1_{13;2}})=(1.3,0.7,1.0,0.5,1.2,0),
$$ 
for the first component and 
$$
\boldsymbol{\beta}|(\psi=2)=(\beta_{0_{12}},\beta_{1_{12}},\beta_{0_{23}}, \beta_{1_{23}},\beta_{0_{13;2}}, \beta_{1_{13;2}})=(1.3,0.2,1.0,0.1,1.2,0),
$$
for the second component.
The comparison between simulated and predicted data shows that the methodology provides accurate density estimation results with asymmetric Archimedean copulas, such as the Clayton and Gumbel.

The results of scenario 4 are illustrated in Figure \ref{DensEst3}, where observations were generated from conditional vines constructed with Clayton copulas and their rotated versions.  
On the left panels we present results from a conditional Clayton vine copula with parameters
$$
\boldsymbol{\beta}=(\beta_{0_{12}},\beta_{1_{12}},\beta_{0_{23}}, \beta_{1_{23}},\beta_{0_{13;2}}, \beta_{1_{13;2}})=(1,0.2,0.8,0.3,0.4,0.1).
$$ 
On the right panels we present results from a conditional vine with rotated Clayton copulas in the first tree and a Clayton copula in the second tree, assuming the following parameters 
$$
\boldsymbol{\beta}=(\beta_{0_{12}},\beta_{1_{12}},\beta_{0_{23}}, \beta_{1_{23}},\beta_{0_{13;2}}, \beta_{1_{13;2}})=(-1.5,-0.4,-0.9,-0.8, 0.4,0.6).
$$
The asymmetric dependence expressed by Clayton and rotated Clayton vines is well captured by the proposed model, as it is clear by comparing simulated and predicted data.

\begin{figure}[htbp]
\centering
     \centering
 \includegraphics[width=1\linewidth]{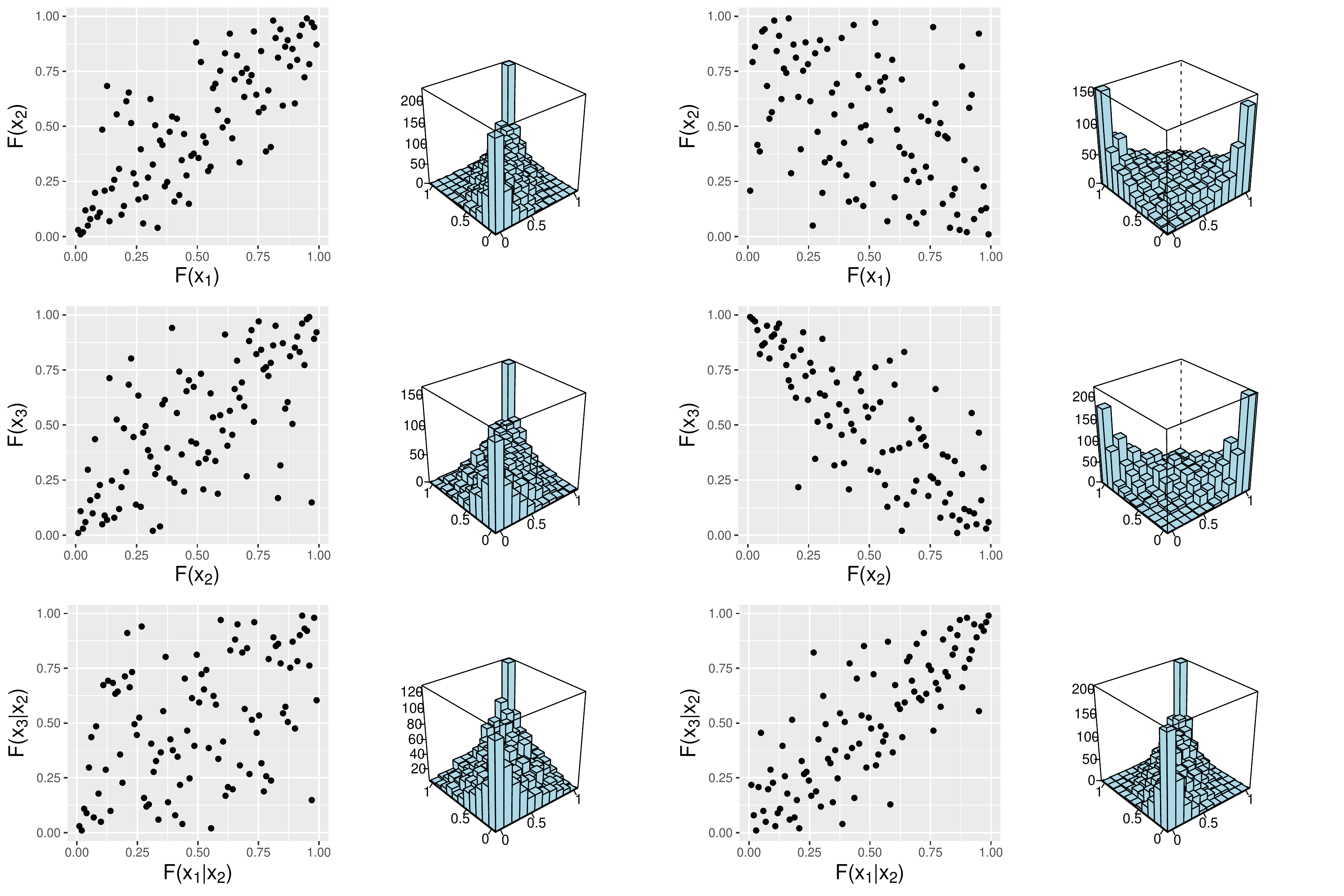}
   \caption{Results from the fifth scenario, where scatterplots show simulated u-data and 3-D histograms show the predicted u-data. The covariate was generated from a N(0.2, 0.1) using the second calibration function.
The left panels depict the results for a conditional Frank vine copula with parameters $\boldsymbol{\beta}=(\beta_{0_{12}},\beta_{1_{12}},\beta_{2_{12}},\beta_{3_{12}},\beta_{0_{23}}, \beta_{1_{23}},\beta_{2_{23}}, \beta_{3_{23}},\beta_{0_{13;2}}, \beta_{1_{13;2}},\beta_{2_{13;2}}, \beta_{4_{13;2}})=(0.7,0.3,0.2,0.1,0.4,0.3,0.1,0.2,0.2,0.4,0.3,0.5)$.
The right panels depict the results for a conditional vine with rotated Gumbel copulas in the first tree and a Gumbel copula in the second tree, with parameters $\boldsymbol{\beta}=(\beta_{0_{12}},\beta_{1_{12}},\beta_{2_{12}},\beta_{3_{12}},\beta_{0_{23}}, \beta_{1_{23}},\beta_{2_{23}}, \beta_{3_{23}},\beta_{0_{13;2}}, \beta_{1_{13;2}},\beta_{2_{13;2}}, \beta_{4_{13;2}})=(-0.3,-0.4,-0.1,0.3,-0.5,-0.6,-0.5,0.8,1,-0.1,0.4,-0.3)$. }
\label{DensEst4}
\end{figure}

In the fifth scenario, the covariate was generated from a N(0.2, 0.1), adopting the second calibration function. 
The results are presented in Figure \ref{DensEst4}, where the scatterplots show the simulated u-data samples, while the 3-D histograms show the predicted u-data samples.
The left panels illustrate the results obtained generating samples from a conditional Frank vine copula with parameters 
\begin{multline*}
\boldsymbol{\beta}=(\beta_{0_{12}},\beta_{1_{12}},\beta_{2_{12}},\beta_{3_{12}},\beta_{0_{23}}, \beta_{1_{23}},\beta_{2_{23}}, \beta_{3_{23}},\beta_{0_{13;2}}, \beta_{1_{13;2}},\beta_{2_{13;2}}, \beta_{4_{13;2}})\\ = (0.7,0.3,0.2,0.1,0.4,0.3,0.1,0.2,0.2,0.4,0.3,0.5).
\end{multline*}
The right panels show the results obtained generating samples from a conditional vine with rotated Gumbel copulas in the first tree and a Gumbel copula in the second three, assuming the following parameters
\begin{multline*}
\boldsymbol{\beta}=(\beta_{0_{12}},\beta_{1_{12}},\beta_{2_{12}},\beta_{3_{12}},\beta_{0_{23}}, \beta_{1_{23}},\beta_{2_{23}}, \beta_{3_{23}},\beta_{0_{13;2}}, \beta_{1_{13;2}},\beta_{2_{13;2}}, \beta_{4_{13;2}}) \\
=(-0.3,-0.4,-0.1,0.3,-0.5,-0.6,-0.5,0.8,1,-0.1,0.4,-0.3).\end{multline*}
The results of Figure \ref{DensEst4} demonstrate the accuracy of our model for density estimation using the second calibration function.  
In all the five considered scenarios, the predictive samples generated from the model parameters in most of the cases replicate perfectly the observations. 

\subsection{Financial Development and Natural Disasters}
We now present the application of the proposed methodology to a heterogeneous dataset to study the impacts of worldwide natural disasters on international financial development.

In recent years several authors in the literature have focused on the relationship between economic growth and natural disasters. \cite{toya2007economic} show that countries with higher income, higher educational attainment, greater openness, more advanced financial systems and smaller governments tend to suffer fewer losses in presence of natural disasters. \cite{felbermayr2014naturally} built a comprehensive database of disaster events and their intensities from primary geophysical and meteorological information, revealing a substantial negative and robust average impact effect of disasters on growth. 
So far, researchers have mostly dealt with financial development studies using several proxies as measures of financial depth, such as the ratio of private credit to GDP or stock market capitalization to GDP. 
For example, \cite{keerthiratne2017impact} proposed an empirical analysis of the effect of the occurrence of natural disasters on financial development proxies.
However, this approach does not take into account the complex nature of financial developments. The International Monetary Fund (IFM) provides, for each country, the Financial Development (FD) index, which summarizes how developed financial institutions and financial markets are in that specific country. The index considers several factors, such as size and liquidity of the markets, ability of individuals and companies to access financial services, ability of institutions to provide financial services at low cost and with sustainable revenues and level of activity of the capital markets. 
We constructed a dataset merging information from the IFM \footnote{Available at \url{https://data.imf.org}.} and the Emergency Events Database \footnote{Available at \url{https://www.emdat.be}.} and 
we analyzed the impact of natural disasters on the dependence between the FD index values in several years. In particular, we investigated how the occurrence of a natural disaster affects the time dependence between the FD index values in the four years following the disaster. 

The original data included observations of the FD index (expressed in percentages) in 181 countries from 1980 to 2019 with annual frequency. We are interested in the effect of a single natural disaster on the FD index in the 4 years following the event. Note that, if in the considered 4 years interval more that one natural disaster occurred, the observation was discarded. 
Based on the original data, we constructed a study dataset where each observation corresponds to a set of 4 consecutive year periods affected by a natural disaster in the first year. Hence, each country may be represented by more than one observation.
Our final study data comprises $N=525$ observations (for $n=86$ different countries) and four variables containing the FD indexes for each one of the 4 considered years. 
The dataset is structured as reported in Table \ref{observations}, where
$t_i$ denotes the number of time periods for each country, with $j = 1, \ldots, t_i$ and $i$ denotes the country, with $i = 1, \ldots, n=86$.

\begin{table}[htbp]
  
  \caption{Financial development data structure.
}
\begin{center}
    \begin{tabular}{|c|c|c|c|c|c|}
    \hline
    Country & Time  & FD index  & FD index  & FD index  & FD index  \\
     &  period & at time 1 &  at time 2 &  at time 3 &  at time 4 \\
    \hline
    1 & 1 & $y_{1,1}$  & $y_{1,1+1}$ & $y_{1,1+2}$ & $y_{1,1+3}$ \\
    1 & 2 & $y_{1,2}$  & $y_{1,2+1}$ & $y_{1,2+2}$ & $y_{1,2+3}$ \\
    \vdots    &  \vdots     &  \vdots     &  \vdots     &  \vdots     &  \vdots \\
   1 & $t_1$    & $y_{1,t_1}$ & $y_{1,t_1+1}$ & $y_{1,t_1+2}$ & $y_{1,t_1+3}$ \\
    \hline
    2 & 1 & $y_{2,1}$  & $y_{2,1+1}$ & $y_{2,1+2}$ & $y_{2,1+3}$ \\
   2 & 2 & $y_{2,2}$  & $y_{2,2+1}$ & $y_{2,2+2}$ & $y_{2,2+3}$ \\
     \vdots     &  \vdots     &  \vdots     &  \vdots     &  \vdots     &  \vdots \\
    2 & $t_2$    & $y_{2,t_2}$ & $y_{2,t_2+1}$ & $y_{2,t_2+2}$ & $y_{2,t_2+3}$ \\
    \hline
     \ldots     &  \ldots     &  \ldots     &  \ldots     &  \ldots     &  \ldots \\
    \hline
    $n$     & 1 & $y_{n,1}$  & $y_{n,1+1}$ & $y_{n,1+2}$ & $y_{n,1+3}$ \\
    $n$     & 2 & $y_{n,2}$  & $y_{n,2+1}$ & $y_{n,2+2}$ & $y_{n,2+3}$ \\
     \vdots     &  \vdots     &  \vdots     &  \vdots     &  \vdots     &  \vdots \\
    $n$     & $t_n$    & $y_{n,t_n}$ & $y_{n,t_n+1}$ & $y_{n,t_n+2}$ & $y_{n,t_n+3}$ \\
    \hline
    \end{tabular}%
\end{center}
  \label{observations}%
\end{table}%

We applied the proposed Bayesian nonparametric unconditional vine mixture approach to the study dataset, constructing a $4$-dimensional vine copula, with marginals denoting the FD index in $4$ consecutive years.
The adoption of an infinite mixture approach allows us to control for both individual and temporal heterogeneity in the dataset.

Since the marginal variables $(y_{i,t_i},y_{i,t_i+1},y_{i,t_i+2},y_{i,t_i+3})$, with $i = 1, \ldots, n$, are expressed in percentages, they are defined on the support $\left [0,1\right ]$. Hence, we assume the marginal distributions to be independent $\text{Beta}(a_{j},b_{j})$ with $a_{j}$ and $b_{j}$ defined a priori $\text{Gamma}(1,1)$ for $j=1,\dots,4$. Inference for the margins is separately performed via Metropolis-Hastings. Results are reported in Table \ref{MarginalsPosterior}.

\begin{table}
\begin{center}
\caption{Financial development and natural disasters analysis: mean, standard deviation and credibility intervals for the posterior densities of the marginal parameters.} 
\label{MarginalsPosterior}

\scalebox{0.8}{
\begin{tabular}{| c |  r r r r r r r r |}
  \hline
 &  $a_{{1}}$ & $b_{{1}}$ &$a_{{2}}$ &  $b_{{2}}$ & $a_{{3}}$ & $b_{{3}}$ &$a_{{4}}$ & $b_{{4}}$ \\
  \hline
$E(\cdot|y)$ & 2.05 & 3.82& 2.17 & 3.85& 2.20 & 3.76 &  2.19 & 3.64 \\ 
 $SD(\cdot|y)$ & 0.11 & 0.23 & 0.13 & 0.25 & 0.13 & 0.23  & 0.12  & 0.21 \\ 
 $q_{0.025}(\cdot|y)$  & 1.83& 3.38& 1.92 & 3.39 & 1.95 &  3.31 &    1.95       & 3.22 \\ 
 $q_{0.975}(\cdot|y)$  & 2.28 & 4.28 & 2.43  & 4.33& 2.45 & 4.21 &   2.44     & 4.08 \\
  \hline 
\end{tabular}
}
\vspace{0.3 cm}
\end{center}
\end{table}

We consider the occurrence of a natural disaster to be a covariate in our conditional vine copula model.
Since natural disasters are defined basing on their intensity, which is measured using the total damage as proxy variable, we constructed the binary covariate taking value $X=1$ if the total damage is over 100 million dollars and $X=0$ otherwise. Since the model requires  a distributional assumption on the covariate, we defined $X \sim \text{Bernoulli}(\phi)$ assuming a priori $\phi\sim\text{Beta}(a_{\phi},b_{\phi})$. Moreover, we chose a linear calibration function for each pair copula $\eta_{s}(x | \boldsymbol{\beta})=\beta_{0_{s}}+\beta_{1_{s}}x$, where $s=1,\dots,6$ denotes the pair-copulas in the vine. 
For the definition of kernel of the mixture we adopt a D-vine approach, since this specific type of vine better describes temporal sequences and easily takes the time ordering of consecutive events into account \citep{barthel2019dependence}. 
We therefore consider a $4$-dimensional D-vine, where the marginals are the FD indexes of the 4 consecutive years and with covariate given by the natural disaster intensity.
We set the total mass parameter as $M=1$ and define the centering measure as 
$$G_0\equiv \text{Beta}\left (a_{\phi},b_{\phi} \right )\times \mathbb{N}_{(6\times 2)}(\boldsymbol{\mu},\boldsymbol{\Sigma}).$$

\begin{figure}
  \centering
    \includegraphics[width=1\linewidth]{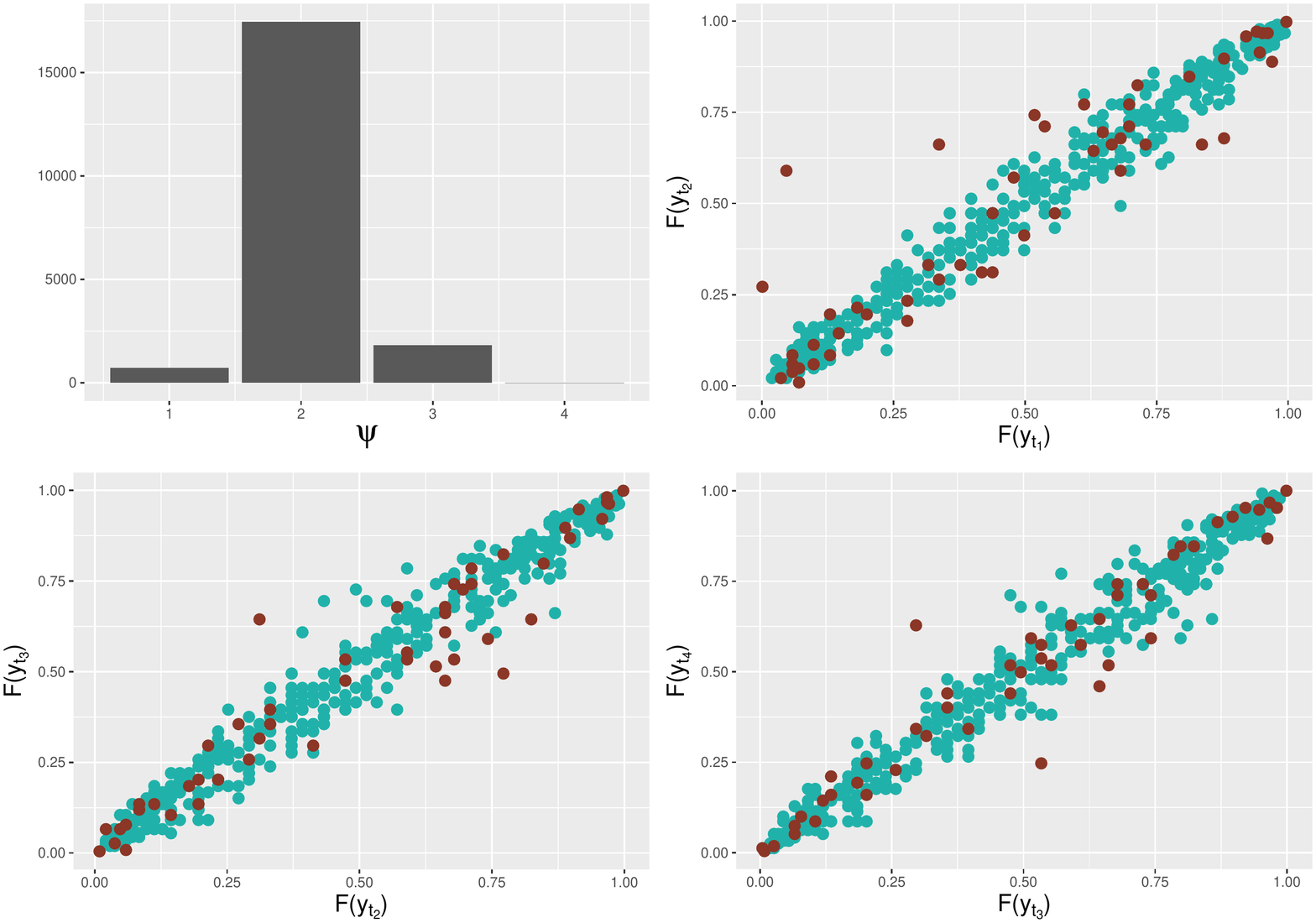}
    \caption{Results from the financial development and natural disasters analysis.
The top left panel shows the barplot of the
mode of the number of observed mixture components. The top right and the
bottom panels show the scatterplots of the observed u-data for the first vine tree; the blue points denote the observations belonging to the first cluster $\psi=1$, while the brown points denote the observations belonging to the second cluster $\psi=2$.}
 \label{FDdata1}
\end{figure}

Our approach estimates two clusters, as shown in Figure \ref{FDdata1}. 
The top left panel shows the barplot of the
mode of the number of observed mixture components. The other panels show the scatterplots of the observed u-data for the first vine tree; 
more precisely, the scatterplot of the first and second time periods are shown in the top right panel, the scatterplot of the second and third time periods are shown in the bottom left panel, while the scatterplot of the third and fourth time periods are shown in the bottom right panel.
The blue points denote the observations belonging to the first cluster $\psi=1$, while the brown points denote the observations belonging to the second cluster $\psi=2$.

In Table \ref{FDdataTab1} we report the resulting mean, standard deviation and credibility intervals for the posterior densities of the calibration functions parameters. The top part of the table shows the results for the first cluster $\psi=1$, while the bottom part shows the results of the second cluster $\psi=2$.

\begin{table}
\begin{center}
\caption{Financial development and natural disasters analysis: mean, standard deviation and credibility intervals for the posterior densities of the calibration function parameters. The top part of the table shows the results for the first cluster $\psi=1$, while the bottom part shows the results of the second cluster $\psi=2$.} 
\label{FDdataTab1}


\scalebox{0.8}{
\begin{tabular}{| c |  r r r r r r r r r r r r   |}
  \hline
$\psi=1$  & $\beta_{0_{12}}$ & $\beta_{1_{12}}$ & $\beta_{0_{23}}$ &$\beta_{1_{23}}$ &  $\beta_{0_{34}}$ & $\beta_{1_{34}}$ & $\beta_{0_{13;2}}$ &$\beta_{1_{13;2}}$ & $\beta_{0_{24;3}}$ & $\beta_{1_{24;3}}$ & $\beta_{0_{14;2,3}}$ & $\beta_{1_{14;2,3}}$ \\
  \hline
$E(\cdot|y,x)$ & 2.40 & -0.63 & 2.31 & -0.49 & 2.36 & -0.81 &               -0.06 & 0.08 &  -0.08 & 0.07 & -0.09 & 0.41 \\ 
 $SD(\cdot|y,x)$ & 0.08 & 0.33 & 0.05 & 0.42 & 0.05 & 0.36  &                 0.06 & 0.33 &  0.06 & 0.26  & 0.06  &0.34 \\ 
 $q_{0.025}(\cdot|y,x)$  & 2.23& -1.34& 2.23 & -1.16 & 2.26 & -1.40&      -0.18       & -0.64& -0.20& -0.41& -0.20& -0.35  \\ 
 $q_{0.975}(\cdot|y,x)$  & 2.51 & -0.05 & 2.39  & 0.62& 2.44 & 0.26&           0.06     & 0.71&  0.04& 0.62& 0.02& 1.03 \\
  \hline 
$\psi=2$  & $\beta_{0_{12}}$ & $\beta_{1_{12}}$ & $\beta_{0_{23}}$ &$\beta_{1_{23}}$ &  $\beta_{0_{34}}$ & $\beta_{1_{34}}$ & $\beta_{0_{13;2}}$ &$\beta_{1_{13;2}}$ & $\beta_{0_{24;3}}$ & $\beta_{1_{24;3}}$ & $\beta_{0_{14;2,3}}$ & $\beta_{1_{14;2,3}}$  \\
  \hline
$E(\cdot|y,x)$ & 0.57 & 1.91 & 1.81 & 0.76 & 1.88 & 0.81 &                       0.01        & -0.19 & -0.08 & 0.14 & -0.13& 0.24  \\ 
 $SD(\cdot|y,x)$ & 0.36 & 0.38 & 0.43 & 0.47 & 0.62 & 0.67 &                    0.26        & 0.32 & 0.60& 0.64&  0.32& 0.40  \\ 
 $q_{0.025}(\cdot|y,x)$  & -0.27 & 1.14 & 0.92 & -0.04 & 0.09& -0.28  &      -0.50       &  -0.50& -0.86& -1.25& -0.73&-0.43   \\ 
 $q_{0.975}(\cdot|y,x)$  & 1.19 & 2.68 & 2.63 & 1.72 & 2.88 & 2.64 &            0.49      & 0.46& 0.44& 0.94& 0.42& 1.03  \\
  \hline 
\end{tabular}
}
\vspace{0.3 cm}
\end{center}
\end{table}

\begin{figure}[htbp]
  \centering
    \includegraphics[width=1\linewidth]{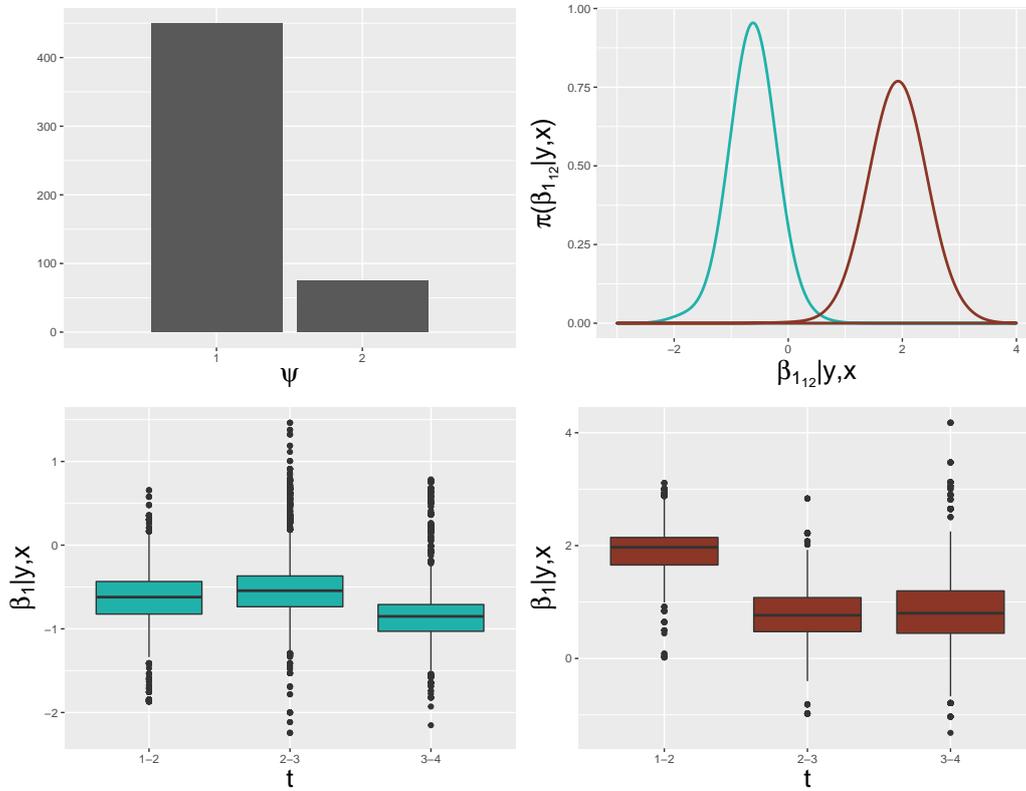}
    \caption{Results from the financial development and natural disasters analysis. The top left panel shows the barplot of the number of observations allocated to the two estimated mixture components. The top right panel compares the posterior densities of the calibration function parameter $\beta_{1_{12}}$ (which is related to the first time interval from $t_i$ to $t_i + 1$, with $i = 1, \ldots, n$) for the first (blue) and the second (brown) mixture components. The left and right bottom panels show, for the first (blue) and second (brown) mixture components, the boxplots of the calibration function parameters $\beta_{1_{12}}$ (left; first time interval from $t_i$ to $t_i + 1$), $\beta_{1_{23}}$ (middle; second time interval from $t_i + 1$ to $t_i + 2$) and $\beta_{1_{34}}$ (right; third time interval from $t_i + 2$ to $t_i + 3$) for the first vine tree.}
 \label{FDdata2}
\end{figure}

The top left panel of Figure \ref{FDdata2} shows the barplot of the number of observations allocated to the two estimated mixture components. The top right panel compares the posterior densities of the calibration function parameter $\beta_{1_{12}}$ (which is related to the first time interval from $t_i$ to $t_i + 1$, with $i = 1, \ldots, n$) for the first (blue) and the second (brown) mixture components. The left and right bottom panels show, for the first (blue) and second (brown) mixture components, the boxplots of the calibration function parameters $\beta_{1_{12}}$ (left; first time interval from $t_i$ to $t_i + 1$), $\beta_{1_{23}}$ (middle; second time interval from $t_i + 1$ to $t_i + 2$) and $\beta_{1_{34}}$ (right; third time interval from $t_i + 2$ to $t_i + 3$) for the first vine tree.
Figure \ref{FDdata2} reveals that the two mixture components present substantial differences in terms of how they are impacted by natural disasters. For the first cluster ($\psi=1$) the model estimates a general negative effect which tends to remain constant until the fourth year;  instead, for the second cluster ($\psi=2$) the model estimates a positive effect of the natural disaster on the time dependence between yearly FD indexes.  

\begin{table}[htbp]
\begin{center}
\caption{Financial development and natural disasters analysis. The left table shows the mean, standard deviation and credibility intervals for the posterior densities of the covariate parameter for the two estimated mixture components ($\psi = 1$ in the left column, $\psi = 2$ in the right column).
The right table shows the same statistics for the mixture weights of the two estimated mixture components ($\psi = 1$ in the left column, $\psi = 2$ in the right column).}
\label{FDdataTab2}
\begin{minipage}[t]{6cm}
\scalebox{0.8}{
\begin{tabular}{| c |  r r   |}
  \hline
 & $\psi=1$ & $\psi=2$  \\
  \hline
$E(\phi|x)$ & 0.03 & 0.84  \\ 
 $SD(\phi|x)$ & 0.01 & 0.07  \\ 
 $q_{0.025}(\phi|x)$  & 0.01 & 0.06  \\ 
 $q_{0.975}(\phi|x)$  & 0.70 & 0.97   \\
  \hline
\end{tabular}
}
\end{minipage}
\hspace{0.2cm}
\begin{minipage}[t]{6cm}
\scalebox{0.8}{
\begin{tabular}{| c |  r r   |}
  \hline
 & $w_1$ & $w_2$  \\
  \hline
$E(\cdot|y,x)$ & 0.85 & 0.15  \\ 
 $SD(\cdot|y,x)$ & 0.02 & 0.02  \\ 
 $q_{0.025}(\cdot|y,x)$  & 0.81 & 0.11  \\ 
 $q_{0.975}(\cdot|y,x)$  & 0.89 & 0.19   \\
  \hline
\end{tabular}
}
\end{minipage}
\end{center}
\end{table}

Let $w_1$ and $w_2$ be the mixture weights for the two estimated components, i.e. the variables indicating the proportion of observations belonging to each group. 
Table \ref{FDdataTab2} shows, on the left, the mean, standard deviation and credibility intervals for the posterior densities of the covariate parameter for the two estimated mixture components ($\psi = 1$ in the left column, $\psi = 2$ in the right column).
On the right, Table \ref{FDdataTab2} shows the same statistics for the mixture weights.
As reported in Table \ref{FDdataTab2}, and in the barplot of Figure \ref{FDdata2}, the $85\%$ of the observations belong to the first cluster, meaning that in most of the cases natural disasters are expected to have negatively affect the financial development. However, from Table \ref{FDdataTab2} we can get another interesting point: the expected value of the posterior probability of natural disasters $E(\phi|x)$ is very low in the first and very high in the second cluster. This result may give a further intuition: the less a natural disaster was expected, the less the government was prepared for that event. On the contrary, if the occurrence of a natural disaster was more expected, governments and financial institutions were prepared for that event and were able to face the possible effects on the financial structure of the country. In Table \ref{group2} we report the list of countries, years and type of natural disasters which had an estimated positive effect on the FD index. 

\begin{table}
\caption{List of countries, years and type of natural disasters which had an estimated positive effect on the FD index.}
\label{group2}
\centering
\scalebox{0.51}{
\begin{tabular}{| l l l  l  |}
  \hline
  Country & $t$ & Category of disaster & Type of disaster \\ 
  \hline
 Burundi & 2015 & Hydrological & Landslide \\ 
 Benin & 1980 & Climatological & Drought \\ 
  Bulgaria & 2007 & Meteorological & Extreme temperature \\ 
Barbados & 1980 & Meteorological & Storm \\ 
   Switzerland & 1994 & Meteorological & Storm \\ 
  Switzerland & 1990 & Meteorological & Storm \\ 
Costa Rica & 2012 & Hydrological & Flood \\ 
 Cyprus & 1995 & Geophysical & Earthquake \\ 
   Denmark & 1999 & Meteorological & Storm \\ 
Denmark & 1992 & Climatological & Drought \\ 
  Estonia & 2005 & Meteorological & Storm \\ 
 Georgia & 2004 & Hydrological & Flood \\ 
 Grenada & 1999 & Meteorological & Storm \\ 
 Croatia & 2000 & Hydrological & Flood \\ 
 Ireland & 2011 & Hydrological & Flood \\ 
Iceland & 2000 & Geophysical & Earthquake \\ 
   Iceland & 1996 & Geophysical & Volcanic activity \\ 
 Israel & 2010 & Hydrological & Flood \\ 
 Israel & 2000 & Meteorological & Extreme temperature \\ 
 Israel & 1992 & Meteorological & Extreme temperature \\ 
 Italy & 2003 & Hydrological & Flood \\ 
 Jamaica & 1980 & Meteorological & Storm \\ 
 Kazakhstan & 2005 & Hydrological & Flood \\ 
 Luxembourg & 1990 & Meteorological & Storm \\ 
 Latvia & 2005 & Meteorological & Storm \\ 
\hline
\end{tabular}
}
\scalebox{0.51}{
\begin{tabular}{| l l l  l |}
  \hline
  Country & $t$ & Category of disaster & Type of disaster \\ 
  \hline
  Madagascar & 1994 & Meteorological & Storm \\ 
 Madagascar & 1986 & Meteorological & Storm \\ 
   Mongolia & 2009 & Hydrological & Flood \\ 
Mongolia & 2003 & Hydrological & Flood \\ 
 Mauritius & 2002 & Meteorological & Storm \\ 
 Norway & 2005 & Meteorological & Storm \\ 
 Norway & 1995 & Hydrological & Flood \\ 
 Norway & 1990 & Meteorological & Storm \\ 
 Panama & 1995 & Hydrological & Flood \\ 
 Poland & 2002 & Meteorological & Storm \\ 
 Poland & 1990 & Meteorological & Storm \\ 
Poland & 1987 & Hydrological & Flood \\ 
Portugal & 2010 & Hydrological & Flood \\ 
 Portugal & 1997 & Meteorological & Storm \\ 
 Portugal & 1985 & Climatological & Wildfire \\ 
 Sweden & 2005 & Meteorological & Storm \\ 
Sweden & 1999 & Meteorological & Storm \\ 
 Sweden & 1990 & Meteorological & Storm \\ 
 Trinidad and Tobago & 1993 & Hydrological & Flood \\ 
 Turkey & 2011 & Geophysical & Earthquake \\ 
 Turkey & 1980 & Hydrological & Landslide \\ 
Ukraine & 2013 & Hydrological & Flood \\ 
Ukraine & 2001 & Hydrological & Flood \\ 
 Uruguay & 2002 & Meteorological & Storm \\ 
  &  &  &  \\ 
   \hline
\end{tabular}
}
\end{table}

\section{Concluding Remarks}

In this paper we introduce a novel approach for multivariate distributions,
which takes advantage of vine specifications.
Vines are very flexible constructions that express multidimensional copulas as building blocks.
This specification allows us to easily incorporate covariates driving the dependence between the response variables.
Inference is carried out following the Bayesian nonparametrics approach, assuming a DP prior distribution on the mixing measure of an infinite mixture of conditional vine copulas.
One of the benefits of this approach is that the specification of copula families for each pair copula is not required.
Another benefit is that this approach allows us to model individual as well as temporal heterogeneity in a natural way.
Our methodology exhibits an excellent performance in simulation studies for both clustering and density estimation.
We also demonstrate the power of the proposed approach applying it to a dataset which investigates the effects of natural disasters on the financial development of a number of countries worldwide.
The proposed Bayesian nonparametric approach for conditional vine mixtures allows us to capture the unobserved heterogeneity that is intrinsic in the dataset.
The results show the presence of two distinct country clusters with opposite behaviour in terms of reaction to natural hazards.



\bibliographystyle{plainnat}
\bibliography{mybib}

\end{document}